\documentclass[aps,twocolumn,showpacs,preprintnumbers,amsmath,notitlepage,nofootinbib,amssymb]{revtex4-1}
\usepackage{amssymb, amsmath,color, hyperref, graphicx}
\usepackage[switch]{lineno}
\usepackage{amsmath,amssymb}
\usepackage[utf8]{inputenc}
\usepackage{graphicx}
\usepackage{mathrsfs}
\usepackage{bm}
\usepackage{bbm}
\usepackage{indentfirst}
\usepackage{epstopdf}
\usepackage{color}
\usepackage{braket}
\usepackage{placeins}
\usepackage{multirow}
\usepackage{lineno}
\usepackage{slashed}
\usepackage{physics}

\newcommand{\pbp}{\langle\bar{\psi}\psi\rangle}

\begin{document}

\title{Chiral condensates and screening masses of neutral pseudoscalar mesons \\
in thermomagnetic QCD medium}

\author{H.-T. Ding}
\address{Key Laboratory of Quark and Lepton Physics (MOE) and Institute of
Particle Physics, Central China Normal University, Wuhan 430079, China}
\author{S.-T. Li}
\affiliation{Key Laboratory of Quark and Lepton Physics (MOE) and Institute of
Particle Physics, Central China Normal University, Wuhan 430079, China}
\author{J.-H. Liu}
\affiliation{Key Laboratory of Quark and Lepton Physics (MOE) and Institute of
Particle Physics, Central China Normal University, Wuhan 430079, China}
\author{X.-D. Wang}
\affiliation{Key Laboratory of Quark and Lepton Physics (MOE) and Institute of
Particle Physics, Central China Normal University, Wuhan 430079, China}

%------------------------------------------------------------------------------------
%       abstract
%------------------------------------------------------------------------------------

%\today
\begin{abstract}
We point out that chiral condensates at nonzero temperature and magnetic fields are in strict connection to the space-time integral of corresponding two-point neutral meson correlation functions in the pseudoscalar channel via the Ward-Takahashi identity. Screening masses of neutral pseudoscalar mesons, which are defined as the exponential decay of the corresponding spatial correlation functions in the long distance, thus are intrinsically connected to (inverse) magnetic catalysis of chiral condensates. To study this we performed lattice simulations of $(2+1)$-flavor QCD on $32^3\times N_t$ lattices with pion mass $M_\pi\simeq 220$ MeV in a fixed scale approach having temperature $T\in[17, 281]$ MeV and magnetic field strength $eB\in[0, 2.5]$ GeV$^2$.
We find that screening lengths, i.e. inverses of screening masses of $\pi^0$, $K^0$ and $\eta^0_{s\bar{s}}$, turn out to have the similar complex $eB$ and $T$ dependences of the corresponding chiral condensates. Although the transition temperature is found to always decrease as $eB$ grows, we show that the suppression due to magnetic fields becomes less significant for hadron screening length and chiral condensates with heavier quarks involved, and ceases to occur for $\eta^0_{s\bar{s}}$ and strange quark chiral condensate. The complex $eB$ and $T$ dependences of both screening masses and chiral condensates, reflecting the crossover nature of the QCD transition, are attributed to the competition between sea and valence quark effects. These findings could be useful to guide low-energy models and effective theories of QCD.
\end{abstract}

%\pacs{11.10.Wx, 11.15.Ha, 12.38.Aw, 12.38.Gc, 12.38.Mh, 24.60.Ky, 25.75.Gz, 25.75.Nq}

\maketitle

%------------------------------------------------------------------------------------
%  introduction
%------------------------------------------------------------------------------------

\section{Introduction} 
\label{sec:intro}
The properties of thermal medium governed by quantum chromodynamics (QCD) in the magnetic fields have attracted a lot of interest recently since the strong magnetic field, which is at the order of QCD scale $\Lambda_{QCD}^2\sim10^4$ MeV$^2\sim10^{17}$ Gauss, is expected to exist in the peripheral relativistic heavy-ion collisions~\cite{Kharzeev:2007jp,Skokov:2009qp,Deng:2012pc,Kharzeev:2020jxw}, the early Universe~\cite{Vachaspati:1991nm}, and magnetars~\cite{duncan1992formation}.
One of the most interesting established features of QCD with physical pions in the strong magnetic field is that the pseudocritical temperature $T_{pc}$ reduces as the magnetic field strength $eB$ grows~\cite{Bali:2011qj}. This, however, was not predicted from all the effective theories/models, since the chiral condensates were previously found to be enhanced by the magnetic fields (so-called magnetic catalyses) at zero temperature and a larger temperature was thus expected to restore the chiral symmetry at nonzero magnetic fields~\cite{Shovkovy:2012zn}. The reduction of $T_{pc}$ accompanies so-called inverse magnetic catalysis, meaning that the averaged up and down quark chiral condensates are suppressed due to $eB$ in the proximity of $T_{pc}$~\cite{Bali:2012zg}.

Many investigations have been taken to understand the unexpected inverse magnetic catalysis as well as the reduction of $T_{pc}$ (see recent reviews, e.g. Refs.~\cite{Andersen:2021lnk,Cao:2021rwx,Bandyopadhyay:2020zte,Andersen:2014xxa}).
%~\red{Common consensus from effective theories/models may need to be added here.} 
It has been established that the inverse magnetic catalysis is driven by the sea quark effects rather than the valence quark effects. In the former the magnetic field is encoded in the fermion matrix but not in the quark propagator, whereas in the latter the magnetic field is only included in the quark propagator~\cite{DElia:2011koc,Bruckmann:2013oba}. However, the connection among the reduction of $T_{pc}$, inverse magnetic catalysis and catalysis  still remains elusive. Moreover, it is found recently that the reduction of $T_{pc}$ in stronger magnetic fields is observed even when the inverse magnetic catalysis of light quark chiral condensates is absent in QCD with very heavy pions~\cite{DElia:2018xwo,Endrodi:2019zrl}. This thus challenges the commonly adopted assumption of strict connection, or even identification between the reduction of $T_{pc}$ and the inverse magnetic catalysis of chiral condensates. 

The reduction of $T_{pc}$ in a background magnetic field, due to the crossover nature of the QCD transition, is also manifested in other observables, e.g. fluctuations of conserved charges~\cite{Ding:2021cwv,Ding:2020pao}, Polyakov loops~\cite{DElia:2018xwo,Bruckmann:2013oba}, the string tension~\cite{Bonati:2016kxj} and the ratio of pressure over energy density~\cite{Bali:2014kia}. These observables may reflect the deconfinement aspect of the QCD transition. On the other hand, the chiral aspect of the QCD transition can be reflected in the properties of light hadrons, e.g. Goldstone pion at zero magnetic fields~\cite{Ding:2015ona,Lahiri:2021lrk}. At $eB=0$ it has been shown that $T_{pc}$ decreases with lighter pions in lattice QCD simulations~\cite{Ding:2019prx,Kotov:2021rah,Ding:2018auz,Li:2020wvy}, and the difference between pion susceptibilities, defined as the integrated pion correlation functions, and chiral susceptibilities, defined as the quark mass derivative of chiral condensates, measures the $SU_L(2)\times SU_R(2)\simeq O(4)$ chiral symmetry breaking~\cite{Ding:2015ona,Bazavov:2019www,Ding:2020xlj,Lombardo:2020bvn}. The pion screening mass, or the inverse of screening length, which is defined as the exponential decay of spatial hadron correlation functions in the long distance, could also reflect the critical behavior near the chiral phase transition~\cite{Rajagopal:1992qz,Son:2001ff}. At zero magnetic fields the pion screening mass in $(2+1)$-flavor QCD was found to increase as either temperature or light quark mass increases~\cite{Brandt:2016daq,Rohrhofer:2019qwq,Dentinger:2021khg,Bazavov:2019www}.

Once the magnetic field is turned on only one Goldstone boson in two-flavor QCD, i.e. $\pi^0$, exists, and the chiral symmetry 
becomes $U_L(1)\times U_R(1)\simeq O(2)$. In analogy to the case at zero magnetic fields the thermal properties of the Goldston boson, $\pi^0$, could be the key to understanding the QCD phase structure, including the reduction of $T_{pc}$ and inverse magnetic catalysis.
Unfortunately, current lattice QCD studies on the hadron properties in nonzero magnetic fields are restricted to the case at zero temperature~\cite{Bali:2011qj,Hidaka:2012mz,Luschevskaya:2014lga,Luschevskaya:2015cko,Bali:2017ian,Bali:2018sey,Endrodi:2019whh,Bignell:2020dze,Ding:2020hxw}. It was found that $\pi^0$ becomes lighter in stronger magnetic fields at zero temperature~\cite{Luschevskaya:2014lga,Luschevskaya:2015cko,Bali:2017ian,Ding:2020hxw}. This, as argued in Refs.~\cite{Bali:2017ian,Ding:2020hxw}, could lead to the reduction of $T_{pc}$ in stronger magnetic fields. On the other hand, studies based on effective theories/models on the pole masses of $\pi^0$ in a background magnetic field in both vacuum and thermal medium have been carried out extensively~\cite{Andersen:2012zc,Fayazbakhsh:2012vr,Kamikado:2013pya,Taya:2014nha,Hattori:2015aki,Andreichikov:2016ayj,Chakraborty:2017vvg,Wang:2017vtn,Li:2020hlp,Ayala:2020dxs,Sheng:2020hge,Avancini:2021pmi,Kojo:2021gvm,Hutauruk:2021dgv,Yang:2021hud,Li:2021swv}. However, similar studies on screening mass of $\pi^0$ in the thermomagnetic medium using effective theories/models are limited~\cite{Sheng:2020hge,Wang:2017vtn,Fayazbakhsh:2012vr,Sheng:2021evj}, and the connection between pole/screening mass of pion and chiral condensates is not established. Moreover, the strange quark could be important in the QCD transition, and its role, however, is not yet studied in the thermomagnetic medium in the current literature.

 In this paper, we point out that, up, down and strange quark chiral condensates are in strict connection to the space-time integral of corresponding two-point neutral meson correlation functions in the pseudoscalar channel. We show the complex $eB$ and temperature dependences of up, down and strange quark chiral condensates and screening masses of $\pi^0$, $K^0$ and $\eta^0_{s\bar{s}}$ extracted from the corresponding spatial correlation functions, as well as the $eB$ dependence of $T_{pc}$. The sea and valence quark effects leading to these phenomena are also discussed.
 
 The paper is organized as follows. In Sec.~\ref{sec:PBP-WI-Msrc} we introduce the Ward-Takahashi identity, and basic observables used for the analyses, e.g. chiral condensates, two-point correlation functions of neutral pseudoscalar mesons as well as screening masses. In Sec.~\ref{sec:setup} we describe our lattice simulations of $N_f=2+1$ QCD using highly improved staggered fermions in a fixed scale approach. In Sec.~\ref{sec:res} we show our main results obtained in the thermomagnetic medium and in Sec.~\ref{sec:summary} we finally draw our conclusions.

\section{Ward-Takahashi identities, chiral condensates and screening masses}
\label{sec:PBP-WI-Msrc}
Although the quark chiral condensates and the neutral pseudoscalar meson correlation function seem to reflect different perspectives of the thermomagnetic medium, they are in strict connection via the following Ward-Takahashi identities at $eB\neq0$, which were derived very recently by some of the current authors~\cite{Ding:2020hxw}:
\begin{align}
\label{eq:WI_pi0}
(m_u+m_d) \,{\chi}_{\pi^0} &= \pbp_u + \pbp_d\,,\\
(m_d+m_s) \,\chi_{K^0} &= \pbp_d + \pbp_s \,, 
\label{eq:WI_K0} \\
m_s \,\chi_{\eta_{s\bar{s}}^0} &= \pbp_s \,.
\label{eq:WI_etas0}
\end{align}
Here $m_{u,d,s}$ is the mass of up, down and strange quark, and $\chi_H$ is the neutral pseudoscalar meson susceptibility and is defined as the integrated Euclidean two-point correlation function of neutral pseudoscalar mesons $H$ with $H=\pi^0$, $K^0$ and $\eta^0_{s\bar{s}}$,
\begin{align}
\chi_{H} = &\int \mathrm{d}z~ G_H(z) \nonumber \\
 \equiv &\int \mathrm{d}z \int_0^{1/T}\mathrm{d}\tau \int \mathrm{d}y\int \mathrm{d}x~\mathcal{G}_H(\tau,\vec{x})\, .
\end{align}
We remark here that the UV divergence appears in the same manner in chiral condensates $\pbp_f$ and meson susceptibilities $\chi_H$.\footnote{$\chi_{H}$ should include contributions from both connected and disconnected diagrams~\cite{Ding:2020hxw}. The contribution from disconnected diagrams arises in, e.g. $\chi_{\pi^0}$ due to isospin symmetry breaking of up and down quarks in nonzero magnetic fields.} The Ward-Takahashi identities shown in Eqs.~\eqref{eq:WI_pi0} and~\eqref{eq:WI_K0} thus naturally bridge the $eB$ dependences of quark chiral condensates, e.g. magnetic catalysis and inverse magnetic catalysis to the properties of Goldstone bosons, $\pi^0$ and $K^0$, in the magnetized medium.\footnote{The Ward-Takahashi identities shown in Eqs.~\eqref{eq:WI_pi0},~\eqref{eq:WI_K0} and~\eqref{eq:WI_etas0} hold true for any values of $T$ and $eB$, and thus the arguments of $T$ and $eB$ are suppressed here to avoid clutter.} Here to check the role of the strange quark we also investigate a fictitious lightest neutral pseudoscalar meson $\eta_{s\bar{s}}^0$ made of a strange and antistrange quark pair.

The single flavor quark chiral condensate $\pbp_f$ at nonzero magnetic fields can be expressed as
\begin{align}
 \pbp_f(B,T) =\frac{T}{V} \frac{\partial \ln Z(B,T)}{\partial m_f}= \frac{T}{V} \mathrm{Tr} M_f^{-1}, 
\label{eq:pbp_q}
\end{align}
where $Z(B,T)$ is the partition function of QCD,
\begin{align}
\label{eq:Z}
    Z(B,T)=\int\mathcal{D}U\,e^{-S_g}\prod_{f=u,d,s}
    \det M_f\,.
\end{align}
Here $V$ is the spatial volume, the Dirac matrix $M_f\equiv M(U,q_fB,m_f)=\slashed{D}(U,q_fB) +m_f\mathbbm{1}$ with $q_f$ and $m_f$ being the electric charge and mass of the quark flavor $f$, and $S_g$ is the gauge action. 

The change of the up and down quark chiral condensates due to the magnetic field is thus reflected in the following renormalized UV-free quantity~\cite{Bali:2012zg}:
\begin{align}
\Delta\Sigma_{ud}(B,T) = \frac{m_u+m_d}{2M_\pi^2 f_\pi^2} 
    \sum_{f=u,d}\left (\pbp_f(B,T) - \pbp_f(0,T)\right )\,.
\label{eq:DSigmaud}
\end{align}
Here $M_\pi$ and $f_\pi$ are the pion  mass and decay constant of pion at zero magnetic field in the vacuum, respectively. These two equations are based on the two flavor version of the Gell-Mann-Oakes-Renner (GMOR) relation~\cite{GellMann:1968rz,Ding:2020hxw}. In the same spirit based on the three flavor GMOR relation~\cite{Gasser:1984gg,Ding:2020hxw}, a quantity involving down and strange quarks can also be defined: 
\begin{align}
\Delta\Sigma_{ds}(B,T)  = \frac{m_d+m_s}{2M_K^2 f_K^2}\sum_{f=d,s}\left (\pbp_f(B,T) - \pbp_f(0,T)\right )\,,
\label{eq:DSigmads}
\end{align}
where $M_{K}$ and $f_{K}$ are the mass and decay constant of the kaon at zero magnetic field in the vacuum, respectively.
One can also investigate the change of strange quark chiral condensate in a similar way,
\begin{align}
\Delta\Sigma_{s}(B,T)  =\frac{m_d+m_s}{2M_{K}^2 f_{K}^2}\left (\pbp_s(B,T) - \pbp_s(0,T)\right )\,.
\label{eq:DSigmas}
\end{align}

On the other hand, two-point Euclidean spatial correlation functions of pseudoscalar mesons are defined as follows\footnote{At zero magnetic field the correlation function is isotropic in the spatial direction, while at nonzero magnetic fields the spatial correlation functions separated along the direction parallel and perpendicular to the magnetic field are different from each other. Examples of such an anisotropy in magnetic fields can also be seen e.g. in the electric conductivity~\cite{Astrakhantsev:2019zkr}, heavy quark potential~\cite{Bonati:2014ksa} and meson deformations~\cite{Hattori:2019ijy}. In our current study we focus on the spatial correlator separated in the direction that is parallel to the direction of the magnetic field, i.e. $G_H(z)$.}:
\begin{align}
     {G}_H(B,T,z)&=\int_0^{1/T}\mathrm{d}\tau \int \mathrm{d}y\int \mathrm{d}x~\mathcal{G}_H(B,\tau,\vec{x}) \nonumber \\
    %  &\sum_{x,y,\tau}\left\langle~{\mathcal{G}}_{f_1f_2}(B,\bm{x})\right\rangle \\=
    &=\frac{\int_0^{1/T}\mathrm{d}\tau \int \mathrm{d}y\int \mathrm{d}x}{Z(B,T)}\int\mathcal{D}U\,e^{-S_g}\\
     &\times \prod_{f=u,d,s} \det M(U,q_fB,m_f)~{\mathcal{G}}_{f_1f_2}(B,\bm{x}), \nonumber
\end{align}
where $\mathcal{G}_{f_1f_2}(B,\bm{x})= \mathcal{O}_{f_1f_2}(B,\bm{x})\left(\mathcal{O}_{f_1f_2}(B,\bm{0})\right)^\dagger$ with $\mathcal{O}_{f_1f_2}(B,\bm{x})=\bar{\psi}_{f_1}(B,\bm{x})
\,\gamma_5\,\psi_{f_2}(B,\bm{x})$ and $\bm{x}\equiv(\tau,\vec{x})\equiv(\tau,x,y,z)$.
The spatial correlator ${G}_H(B,T,z)$ decays exponentially at a large spatial distance $z$~\cite{Detar:1987hib,Detar:1987kae}:
\begin{equation}
\lim_{z \rightarrow\infty} G_H(B,T,z) = A_H~ e^{-M_Hz},
\label{eq:defMscr}
\end{equation}
where $A_H$ is the amplitude and $M_H$ is the screening mass or the inverse of the screening length of the corresponding hadron $H$. At zero temperature the screening mass $M_H$ is the same as the pole mass of the ground state of pseudoscalar meson $H$, and at nonzero temperature they differ among each other~\cite{Karsch:2003jg}.

To investigate the sea and valence quark effects to the behavior of the screening masses, we thus define the following two types of spatial correlation functions following the similar techniques used for light quark chiral condensates in Refs.~\cite{DElia:2011koc,Bruckmann:2013oba}:
\begin{align}
\label{eq:val-corr}
     {G}^{\rm val}_H(B,T,z)&=\frac{\int_0^{1/T}\mathrm{d}\tau \int \mathrm{d}y\int \mathrm{d}x}{Z(B=0,T)}\int\mathcal{D}U\,e^{-S_g}\\
     &\times \prod_{f=u,d,s} \det M(U,q_fB=0,m_f)~{\mathcal{G}}_{f_1f_2}(B,\bm{x}), \nonumber  \\
          \label{eq:sea-corr}
    {G}^{\rm sea}_H(B,T,z)&=\frac{\int_0^{1/T}\mathrm{d}\tau \int \mathrm{d}y\int \mathrm{d}x}{Z(B,T)}\int\mathcal{D}U\,e^{-S_g}\\
     &\times \prod_{f=u,d,s} \det M(U,q_fB,m_f)~{\mathcal{G}}_{f_1f_2}(B=0,\bm{x}). \nonumber 
\end{align}
Thus the $eB$ dependence of $G^{\rm val}_H(B,T,z)$ only comes from ${\mathcal{G}}_{f_1f_2}(B,\bm{x})$, whereas the $eB$ dependence of $G^{\rm sea}_H(B,T,z)$ only comes from the fermion matrix $M(U,q_fB,m_f)$. The screening masses $M^{\rm val}_H$ and $M^{\rm sea}_H$ can thus be extracted from $G^{\rm val}_H(B,T,z)$ and $G^{\rm sea}_H(B,T,z)$ according to Eq.~\eqref{eq:defMscr}, respectively.

\section{Lattice QCD calculations}
\label{sec:setup}
The highly improved staggered quarks (HISQ)~\cite{Follana:2006rc} and a tree-level improved Symanzik gauge action, which have been extensively used by the HotQCD collaboration~\cite{Bazavov:2011nk, Bazavov:2014pvz, Bazavov:2012jq, Bazavov:2017dus,
Bazavov:2018mes}, were adopted in our current lattice simulations of $N_f=2+1$ QCD in nonzero magnetic fields. The magnetic field described by a complex phase factor is introduced along the $z$ direction, and is implemented by multiplying the gauge links of lattices. To satisfy the quantization for all the quarks in the system the magnetic field strength $eB$ is thus expressed as follows~\cite{Bali:2011qj,DElia:2010abb}:
\begin{equation}
eB= \frac{6 \pi N_{b}}{N_{s}^2} a^{-2} \equiv 6 \pi N_{b} T^2\frac{N_t^2}{N_{s}^2},
\label{eq:eBdef}
\end{equation}
where $N_b \in \mathbf{Z}$ is the number of magnetic fluxes through a unit area in the $x$-$y$ plane, and $a$ is the lattice spacing. Here $N_s$ and $N_t$ denote the number of points in the spatial and temporal direction, respectively.
The implementation of magnetic fields in the lattice QCD simulations using the HISQ action is detailed in Ref.~\cite{Ding:2020hxw}.

In our lattice simulations, the strange quark mass is fixed to its physical value $m_{s}^{\rm phy}$ by tuning the mass of a (fictitious) $s\bar{s}$ pseudoscalar meson, $\eta^0_{s\bar{s}}$, to $M_{\eta^0_{s\bar{s}}}\simeq~684$ MeV~\cite{Ding:2020hxw}.
The contribution from the disconnected diagram to the mass of $\eta^0_{s\bar{s}}$ is generally not considered and the mass of $\eta^0_{s\bar{s}}$ can also be
estimated using leading order chiral perturbation theory $M_{\eta^0_{s\bar{s}}}=\sqrt{2M_{K}^2-M_{\pi}^2}$~\cite{Bazavov:2014cta,Bazavov:2019www}. The light quark mass is chosen as $m_u=m_d=m_s^{\rm phy}$/10 and this corresponds to a Goldstone pion mass $M_{\pi }\simeq220$ MeV, kaon mass $M_{K}\simeq~$507 MeV. In the current lattice setup pion and kaon decay constants are $f_\pi\simeq97$ MeV and $f_K\simeq113$ MeV in the vacuum~\cite{Ding:2020hxw}, and the pseudocritical temperature $T_{pc}(eB=0)\approx 170$ MeV as estimated from the disconnected light quark chiral susceptibility~\cite{Ding:2021cwv}.

Due to the different electric charge of up and down (strange) quarks they are treated differently in nonzero magnetic fields. To have the same value of $eB$ in physical units at different temperatures we adopted a fixed scale approach, i.e. fixed lattice spacing $a\simeq0.117$ fm in our simulations. We fix $N_s=32$ and vary $N_t$ from 96 to 6 in order to have 8 different temperatures $T=(aN_t)^{-1}$ ranging from 17 MeV up to 281 MeV. For each fixed $N_{t}$, we have 15 values of magnetic field flux $N_{b}$ chosen from 0 to 48. These correspond to the magnetic field $eB$ ranging from 0 to $\sim$2.5 GeV$^2$. The gauge field configurations were generated using the rational hybrid Monte Carlo
algorithm~\cite{Clark:2004cp,Bazavov:2010ru}. An average of about 3000 gauge configurations separated by every 5$\mathrm{th}$ molecular dynamics trajectory of unit length were saved for each parameter set to carry out
various measurements. More information on these configurations can be found in~\cite{Ding:2020hxw,Ding:2021cwv}.

Measurements of chiral condensates were done by inverting the fermion matrices using 102 Gaussian random sources while those of correlation functions were done using a single corner wall source on each gauge configuration. It has been shown in Ref.~\cite{Luschevskaya:2015cko} that the quark-line disconnected part in quenched QCD is negligible to the neutral pion correlation function in the nonzero magnetic fields. In our current study of neutral pions, we thus neglect the disconnected contributions which could be small as well~\cite{Ding:2020hxw}. To quantify the sea and valence quark effects we compute $G^{\rm sea}_H(B,z)$ based on the gauge configurations produced at $eB\neq0$ and with ${\mathcal{G}}_{f_1f_2}(B=0,\bm{x})$, and compute $G^{\rm val}_H(B,z)$ based on the gauge configurations produced at $eB=0$ and with ${\mathcal{G}}_{f_1f_2}(B\neq0,\bm{x})$. Detailed information on the techniques adopted in the current work to compute  chiral condensates, correlation functions and screening masses in the staggered theory can be found in e.g. Sec. III B in both Refs.~\cite{Bazavov:2019www,Ding:2020hxw}. A typical example of extraction of screening masses at $T~$=~140 MeV is also shown in Appendix~\ref{sec:app_method}.

\section{Results}
\label{sec:res}
\begin{figure}[!htbp]
	\centering
\includegraphics[width=0.45\textwidth]{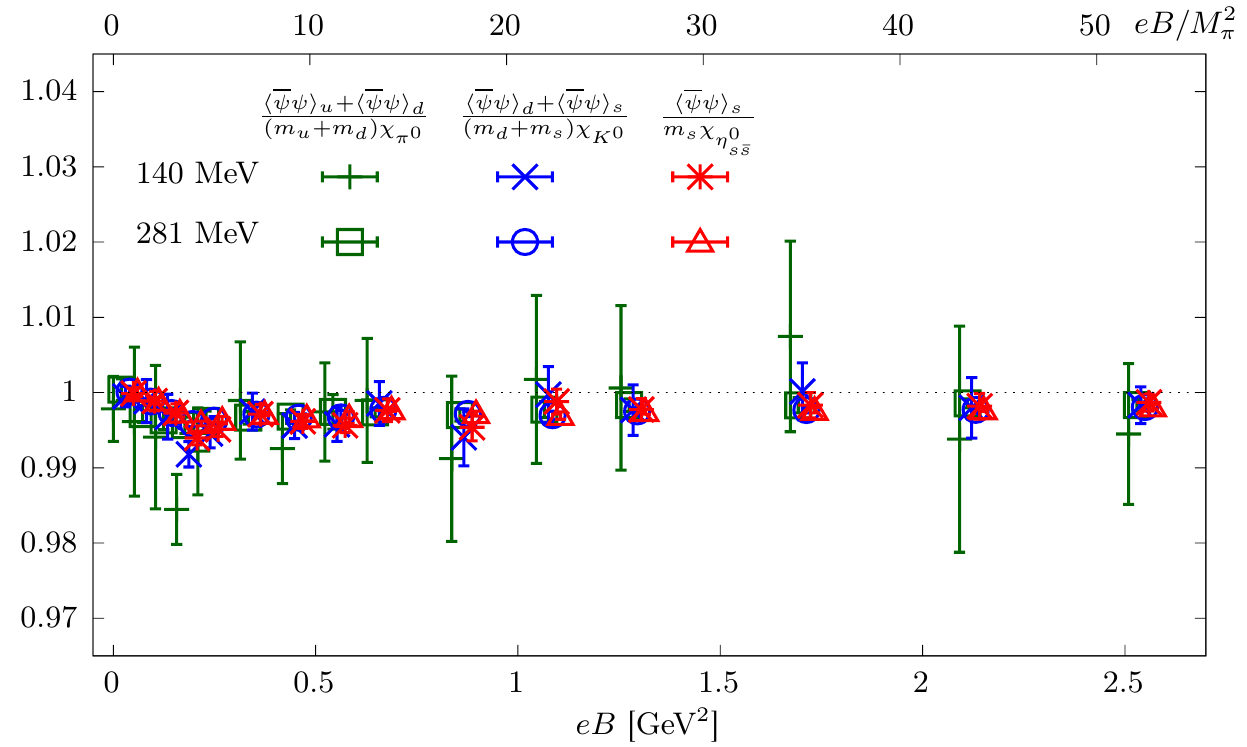}
	\caption{Ratios of $(\pbp_u+\pbp_d)/((m_u+m_d)\chi_{\pi^0})$, $(\pbp_d+\pbp_s)/((m_d+m_s)\chi_{K^0})$ and $\pbp_s/(m_s\chi_{\eta^0_{s\bar{s}}})$ as function of $eB$ at $T=140$ and 281 MeV. Data points are shifted horizontally for visibility.}
	\label{fig:WI}
\end{figure}
To demonstrate the Ward-Takahashi identities [cf. Eqs.~\eqref{eq:WI_pi0}, \eqref{eq:WI_K0} and \eqref{eq:WI_etas0}] we show the ratios of $(\pbp_u+\pbp_d)/((m_u+m_d)\chi_{\pi^0})$, $(\pbp_d+\pbp_s)/((m_d+m_s)\chi_{K^0})$ and $\pbp_s/(m_s\chi_{\eta^0_{s\bar{s}}})$  as a function of $eB$ at two example temperatures in Fig.~\ref{fig:WI}. It can be read off from the plot that the ratios are consistent with unity according to the Ward-Takahashi identities.\footnote{Note that the deviation of $(\pbp_u+\pbp_d)/((m_u+m_d)\chi_{\pi^0})$ from unity, i.e. at most $\sim$2 percent, might originate from the contribution of the disconnected diagrams to $\chi_{\pi^0}$ that is neglected in the current study.} The case is the same at other temperatures, see e.g. $T~$=~17 MeV shown in Ref.~\cite{Ding:2020hxw}. This thus shows the strict connection between chiral condensates and neutral pseudoscalar meson correlation functions in the thermomagnetic medium.

\begin{figure}[!thbp]
	\centering
	\includegraphics[width=0.45\textwidth]{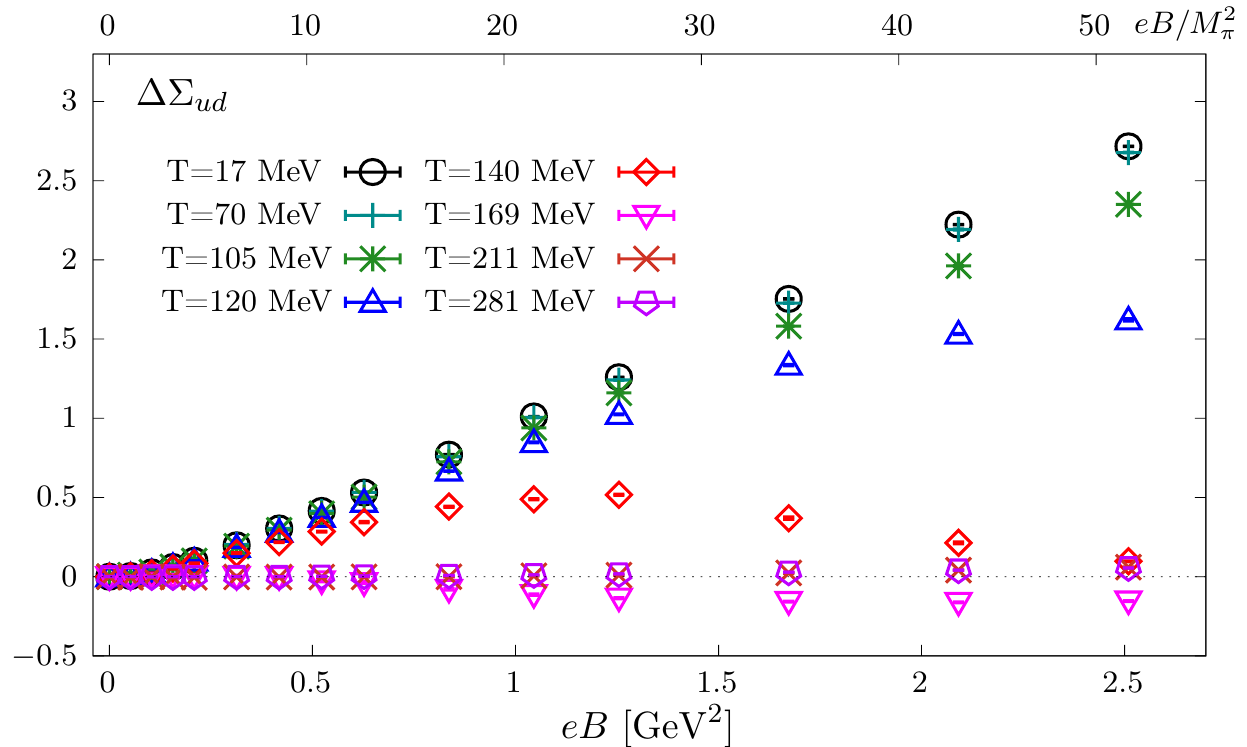}
		\includegraphics[width=0.45\textwidth]{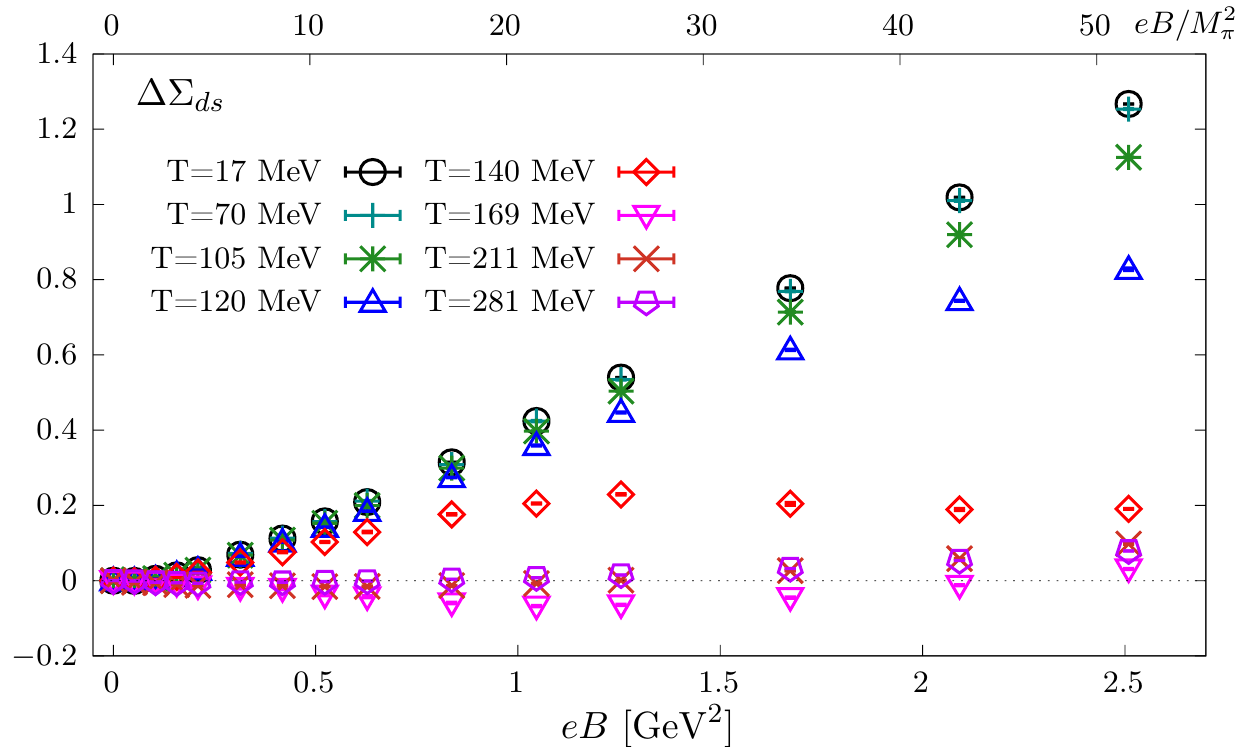}
	\includegraphics[width=0.45\textwidth]{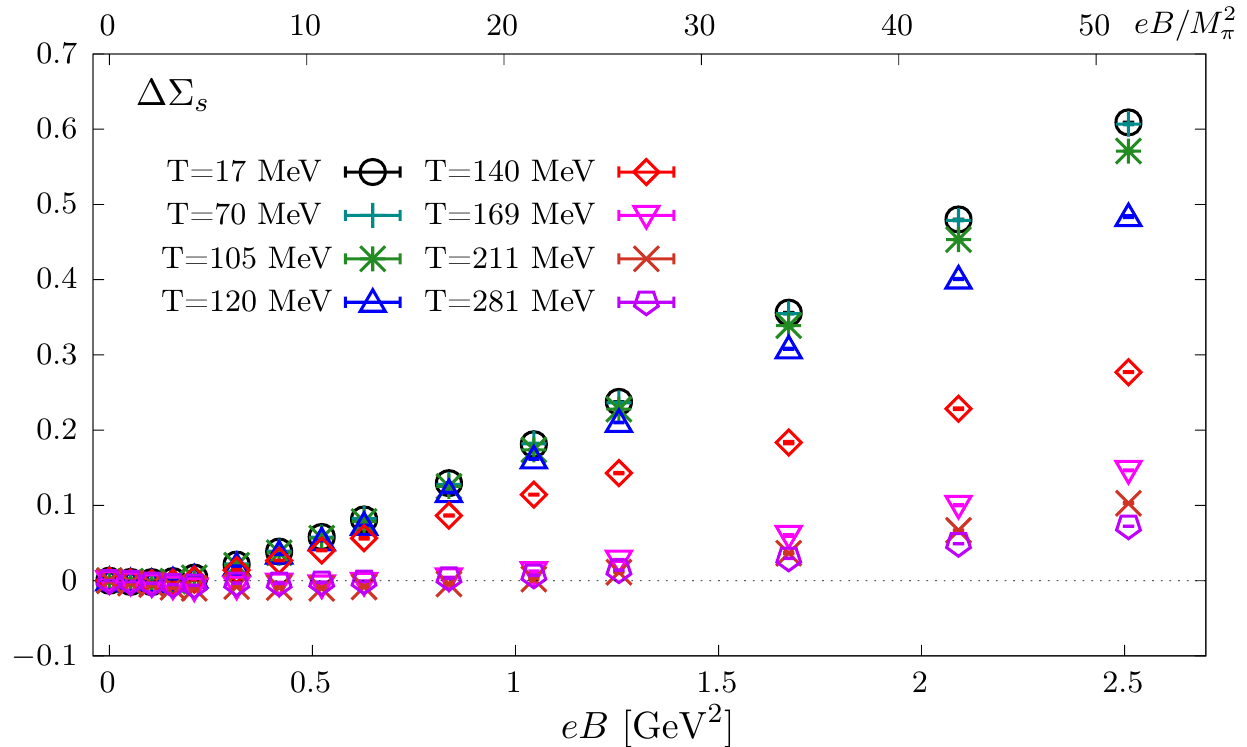}
	\caption{The change of the renormalized up and down quark chiral condensates, $\Delta\Sigma_{ud}$ (top), down and strange quark chiral condensates, $\Delta\Sigma_{ds}$ (middle), strange quark chiral condensate $\Delta\Sigma_{s}$ (bottom) due to the magnetic field as a function of $eB$ at various temperatures. The upper $x$ axis in each plot is rescaled by the pion mass square in the vacuum at $eB=0$. }
	\label{fig:PBP-eB}
\end{figure}

In Fig.~\ref{fig:PBP-eB} (top) we show the change of up and down quark chiral condensates due to $eB$, $\Delta\Sigma_{ud}$, as a function of $eB$ with temperatures from $T=17$ MeV up to $T=281$ MeV. The magnetic catalysis of chiral condensates can be clearly observed at $T\leq 120$ MeV, while the inverse magnetic catalysis can be seen at $T=140$ and 169 MeV in the proximity of $T_{pc}(eB=0)\simeq170$ MeV. At the highest two temperatures $\Delta\Sigma_{ud}$ are consistent with zero. These observations are in line with findings from lattice QCD studies with physical pion mass~\cite{Bali:2012zg}.

\begin{figure}[!htbp]
	\centering
		\includegraphics[width=0.45\textwidth]{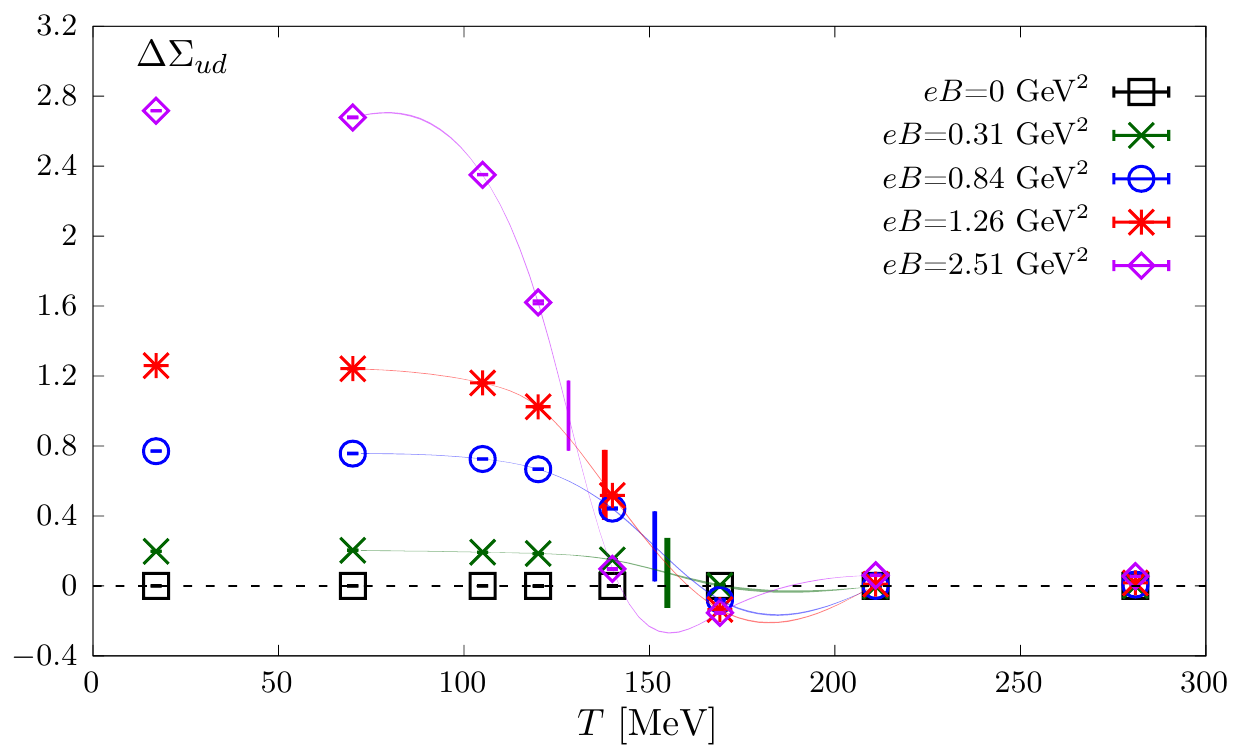}
		\includegraphics[width=0.45\textwidth]{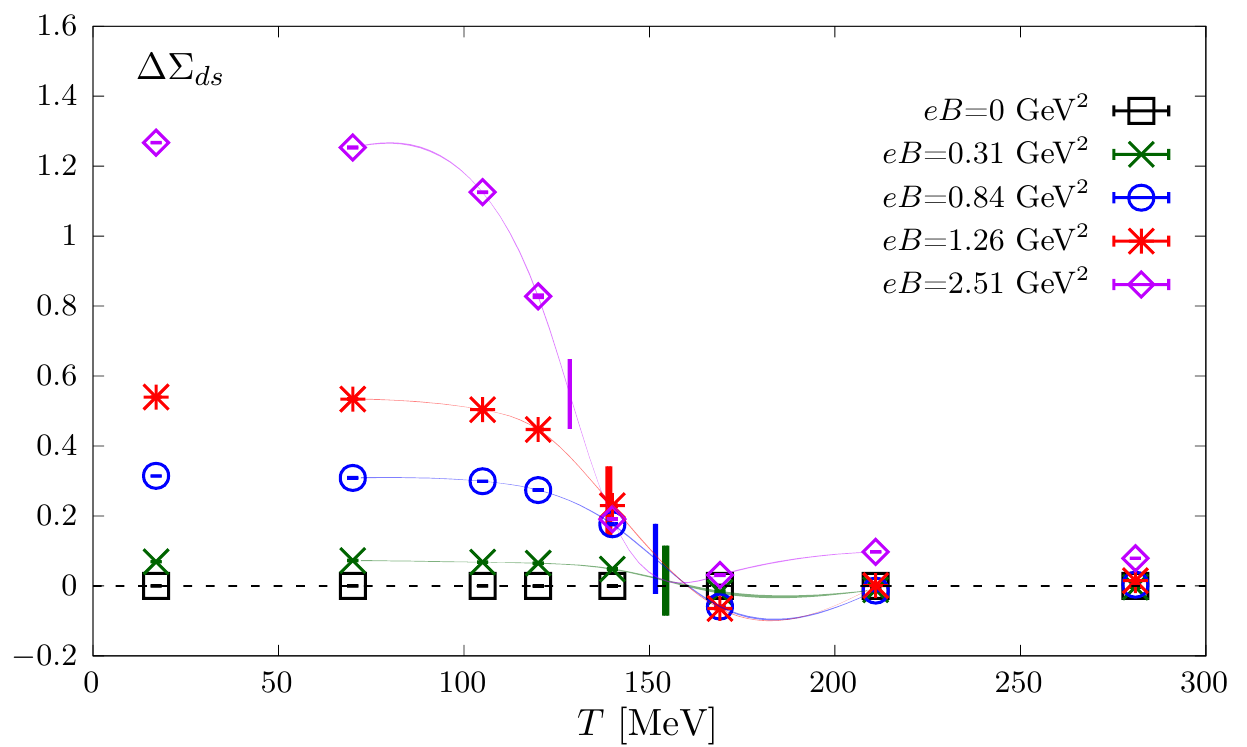}
	\includegraphics[width=0.45\textwidth]{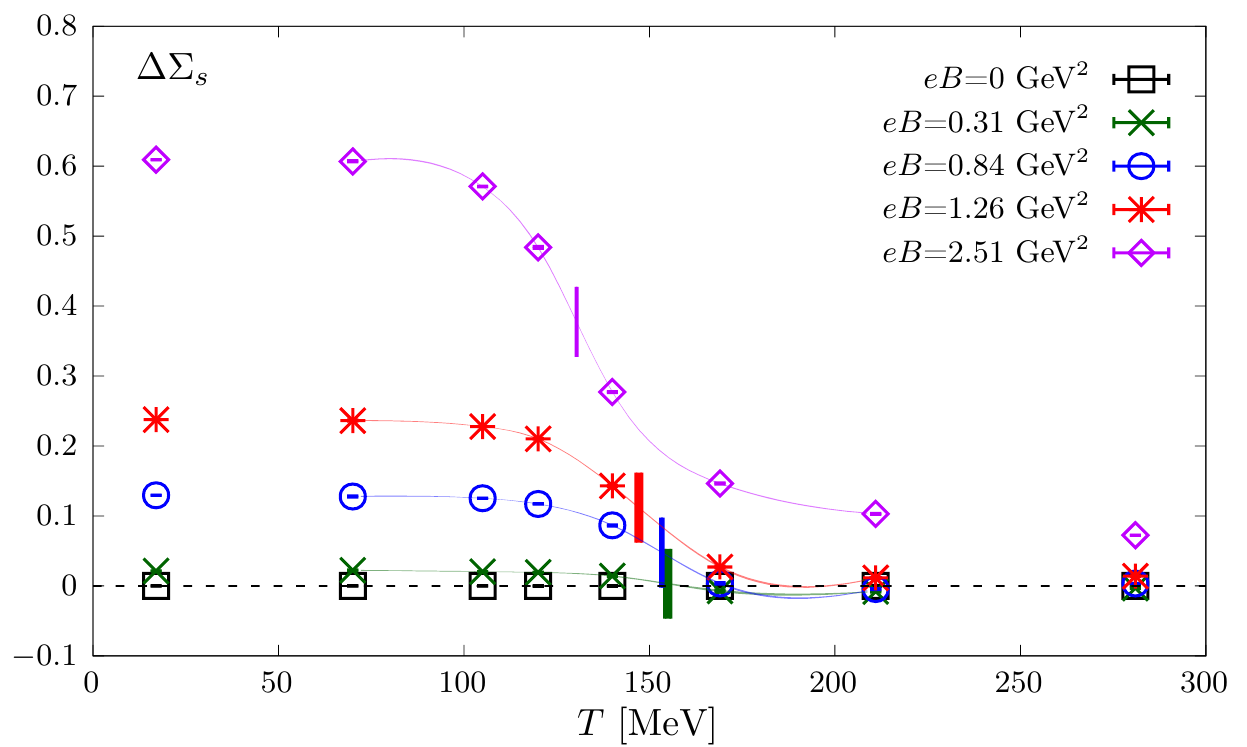}
	\caption{$\Delta\Sigma_{ud}$ (top), $\Delta\Sigma_{ds}$ (middle) and $\Delta\Sigma_{s}$ (bottom) as a function of $T$ at several values of magnetic field strength $eB$. Bands connecting the data points represent the spline interpolations, while the center and half width of each rectangle show the location and uncertainty of the inflection point, respectively.}
	\label{fig:PBP-T}
\end{figure}

In Fig.~\ref{fig:PBP-eB} (middle) we show a similar plot as $\Delta\Sigma_{ud}$ but for the change of down and strange quark condensates due to $eB$, $\Delta\Sigma_{ds}$. At $T\leq140$ MeV $\Delta\Sigma_{ds}$ has similar $eB$ dependences as $\Delta\Sigma_{ud}$, i.e. an increasing behavior of $\Delta\Sigma_{ds}$ at $T\leq120$ MeV, and a nonmonotonous behavior of $\Delta\Sigma_{ds}$ although less significant compared to $\Delta\Sigma_{ud}$ at $T=140$ MeV. At higher temperatures, i.e. $T\geq169$ MeV $\Delta\Sigma_{ds}$ instead shows an increasing trend as $eB$ grows with $eB\gtrsim$ 1.26 GeV$^2$. This increasing trend in $\Delta\Sigma_{ds}$ comes from the strange quark chiral condensate (Fig.~\ref{fig:PBP-eB} bottom). At $T=169$ MeV $\Delta\Sigma_{s}$ shows a marked increasing behavior in $eB$. Furthermore, $\Delta\Sigma_{s}$ is found to always increase as $eB$ grows in the current window of $eB$ and $T$. In other words, only magnetic catalysis is found in $\Delta\Sigma_{s}$. This is in contrast to both $\Delta\Sigma_{ud}$ and $\Delta\Sigma_{ds}$.

In Fig.~\ref{fig:PBP-T} we show $\Delta\Sigma_{ud}$ (top), $\Delta\Sigma_{ds}$ (middle) and $\Delta\Sigma_{s}$ (bottom) as a function of temperature at several values of $eB$. It can be seen that the QCD transition in the current window of the magnetic field is a rapid crossover and the strength of the transition becomes stronger as $eB$ grows. Due to the crossover nature of the QCD transition, the inflection points of all these quantities as a function of temperature can be used to define the pseudocritical temperature $T_{pc}$ of QCD. It can be clearly observed that the inflection points of all these quantities shift to lower temperatures in stronger magnetic fields. This thus means that $T_{pc}$ obtained from these inflections always decreases as $eB$ grows. We remark here that although $\Delta\Sigma_{s}$ does not show any marked decreasing behavior in $eB$ at each fixed temperature [cf. Fig.~\ref{fig:PBP-eB} (bottom)] its inflection point still moves to lower temperatures in stronger magnetic fields similar to the case of $\Delta\Sigma_{ud}$ and $\Delta\Sigma_{ds}$. It is also worth mentioning that Fig.~\ref{fig:PBP-T} is in analogy to Fig. 14 in Ref.~\cite{DElia:2018xwo}.\footnote{Figure 6 in Ref.~\cite{Endrodi:2019zrl} shows similar results as Fig. 14 in Ref.~\cite{DElia:2018xwo} but along the transition line of $T_{pc}(m_{u,d})$.} Both figures show that $T_{pc}(eB)$ decreases as $eB$ grows even when the relevant observable is no longer suppressed due to the magnetic field. However, in the former case the relevant observable is the strange quark chiral condensate in QCD with $M_\pi(eB=0,T=0)$ fixed to 220 MeV, whereas in the latter case the observable is the light quark chiral condensates with $M_\pi(eB=0,T=0)=664$ MeV~\cite{DElia:2018xwo}.

\begin{figure}[!htbp]
	\centering
	\includegraphics[width=0.45\textwidth]{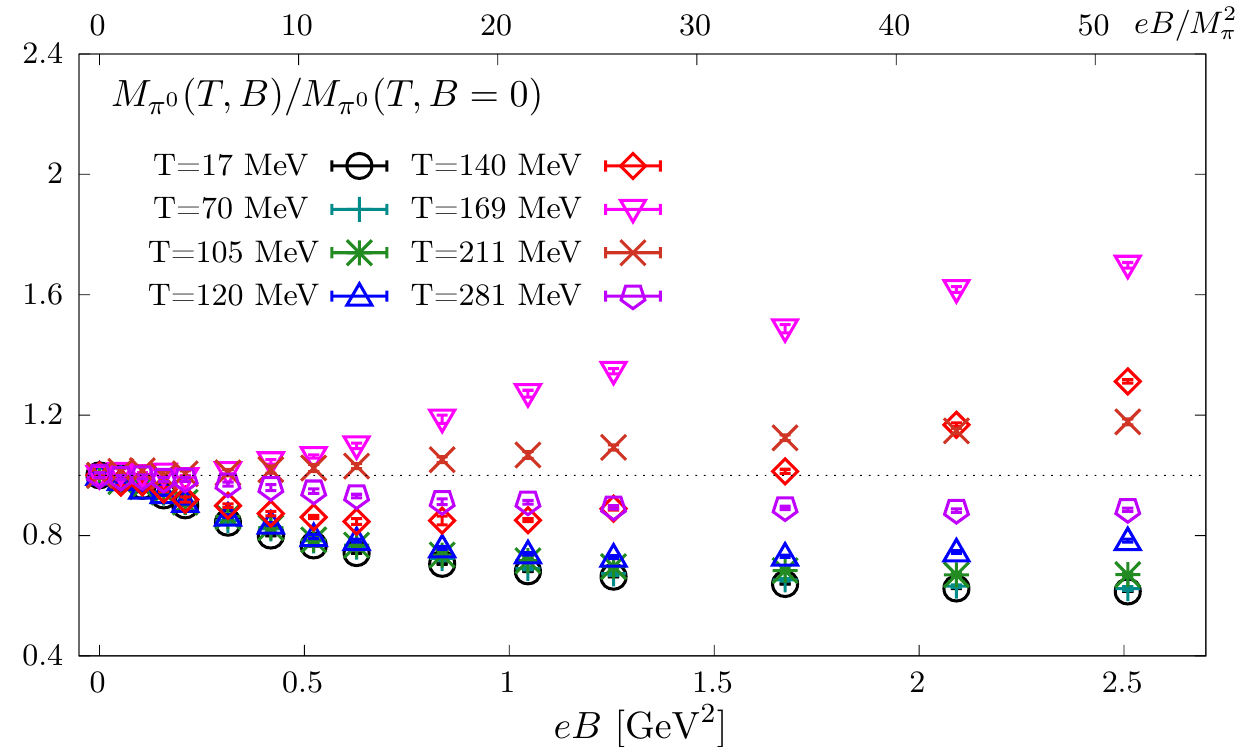}
		\includegraphics[width=0.45\textwidth]{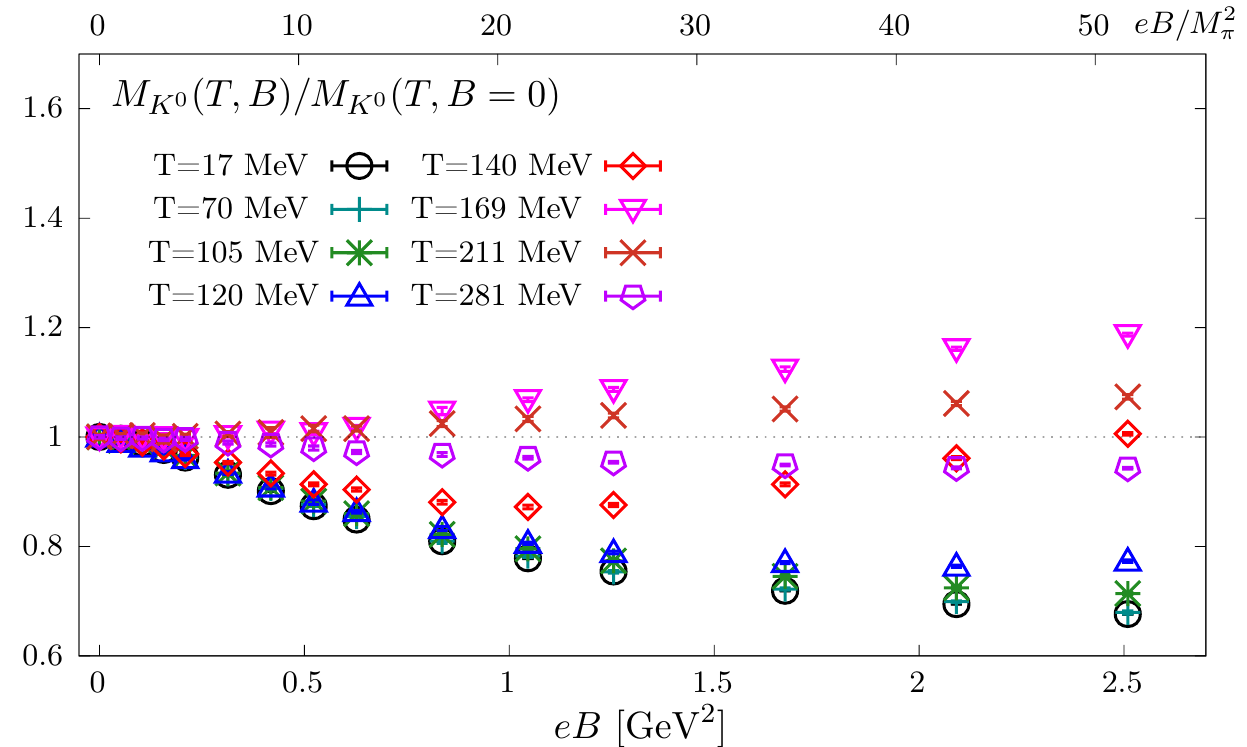}
	\includegraphics[width=0.45\textwidth]{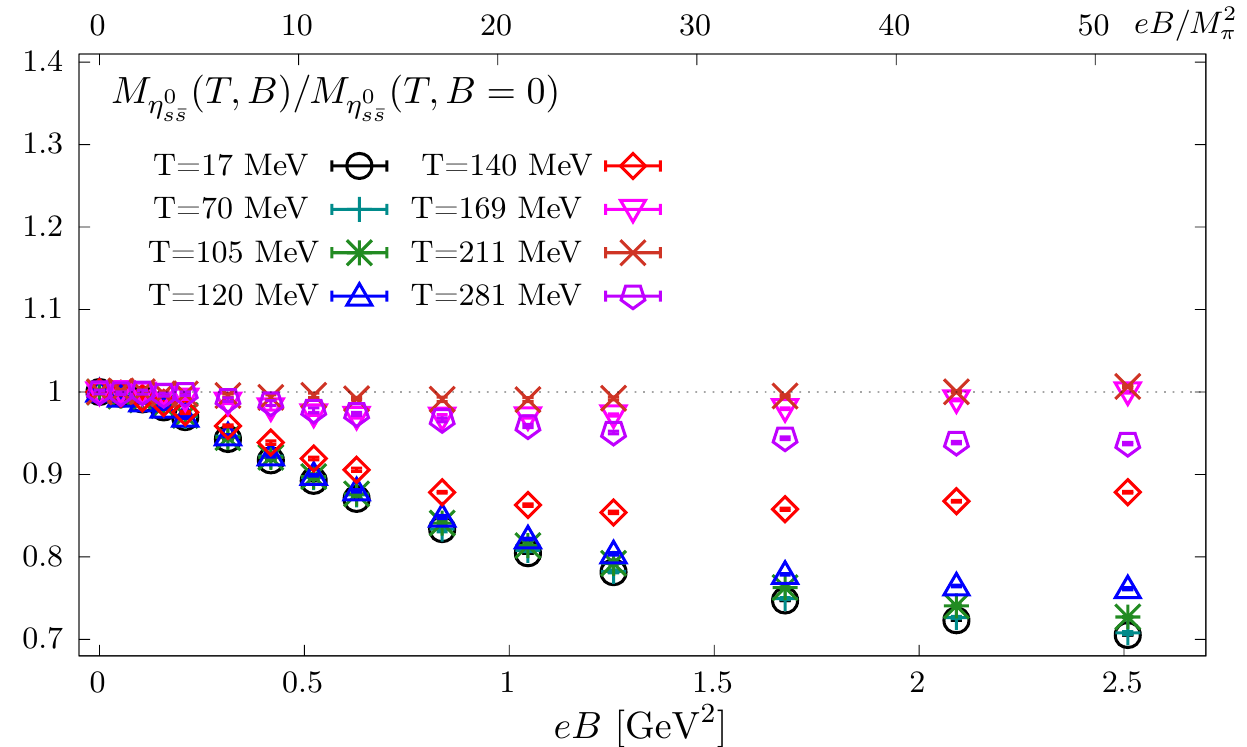}
	\caption{Ratios of screening masses of $\pi^0$ (top), $K^0$ (middle) and $\eta^0_{s\bar{s}}$ (bottom) to their corresponding values at $eB=0$ as function of $eB$ at various temperatures.}
	\label{fig:Mscr-eB}
\end{figure}

In Fig.~\ref{fig:Mscr-eB} (top) we show the ratios of screening masses to their values at $eB=0$ and same temperature for $\pi^0$ extracted from the corresponding spatial correlators. Although $\chi_{\pi^0}$ or $\Delta\Sigma_{ud}$ is enhanced at low temperature and suppressed around the transition temperature due to the magnetic field, it is not fully expected that the screening length, or the inverse of the screening mass of the neutral pion follows the same $eB$ and temperature dependence of $\chi_{\pi^0}$. This is because it is the long distance part of the correlation function that decays exponentially with the screening mass $M_{\pi^0}$ and $A_{\pi^0}$ should also depend on $eB$ [cf. Eq.~\eqref{eq:defMscr}]. It turns out that the extracted screening length of $\pi^0$ has similar $eB$ dependences of $\Delta\Sigma_{ud}$ and $\chi_{\pi^0}$ at each fixed temperature. At $T$~=~17 MeV, i.e. much lower to the transition temperature, the screening mass is the same as the pole mass and decreases as $eB$ grows. As discussed in Ref.~\cite{Ding:2020hxw} the neutral pion cannot be considered as a pointlike particle anymore in the current window of the magnetic fields since a neutral pointlike particle should be blind to the magnetic field. At higher temperatures but $T~\leq 105$ MeV the neutral pion screening mass $M_{\pi^0}$ still decreases as $eB$ grows. With increasing temperature, i.e. at $T$~=~120 and 140 MeV $M_{\pi^0}$ first decreases and then starts to increase after a turning point at $eB\simeq 1.26$ GeV$^2$. At higher temperatures, i.e. $T$~=~169 and 211 MeV, $M_{\pi^0}$ show marked increasing behavior with increasing $eB$, whereas at the highest temperature $T$~=~281 MeV, $M_{\pi^0}$ shows a decreasing behavior again. It is also worth mentioning that  the ratio $M_\pi^0(T,B)/M_\pi^0(T,B=0)$ is always smaller than unity at $T\leq120$ MeV and $T$~=~281 MeV, and larger than unity at $T$~=~169 and 211 MeV, while at $T$~=~140 MeV it is smaller than unity with $eB<$ 2 GeV$^2$ and becomes larger than unity with $eB\gtrsim$ 2 GeV$^2$.

\begin{figure}
	\centering
	\includegraphics[width=0.45\textwidth]{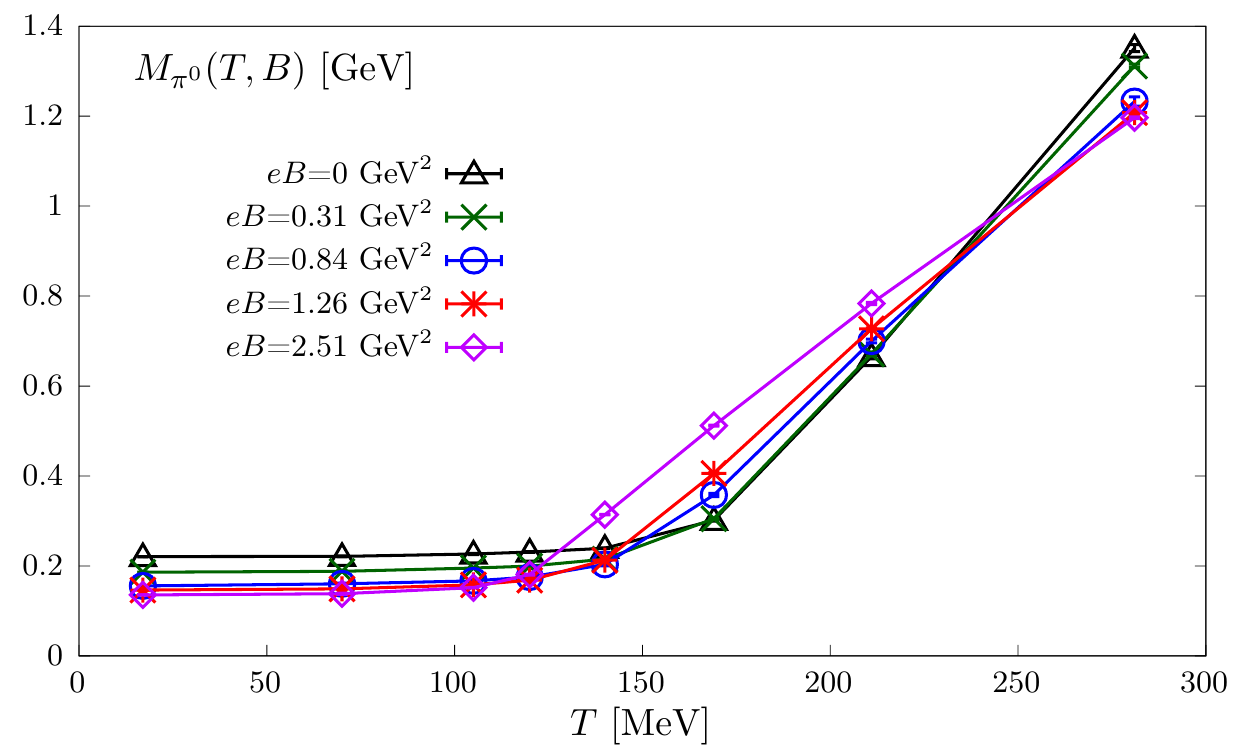}
\includegraphics[width=0.45\textwidth]{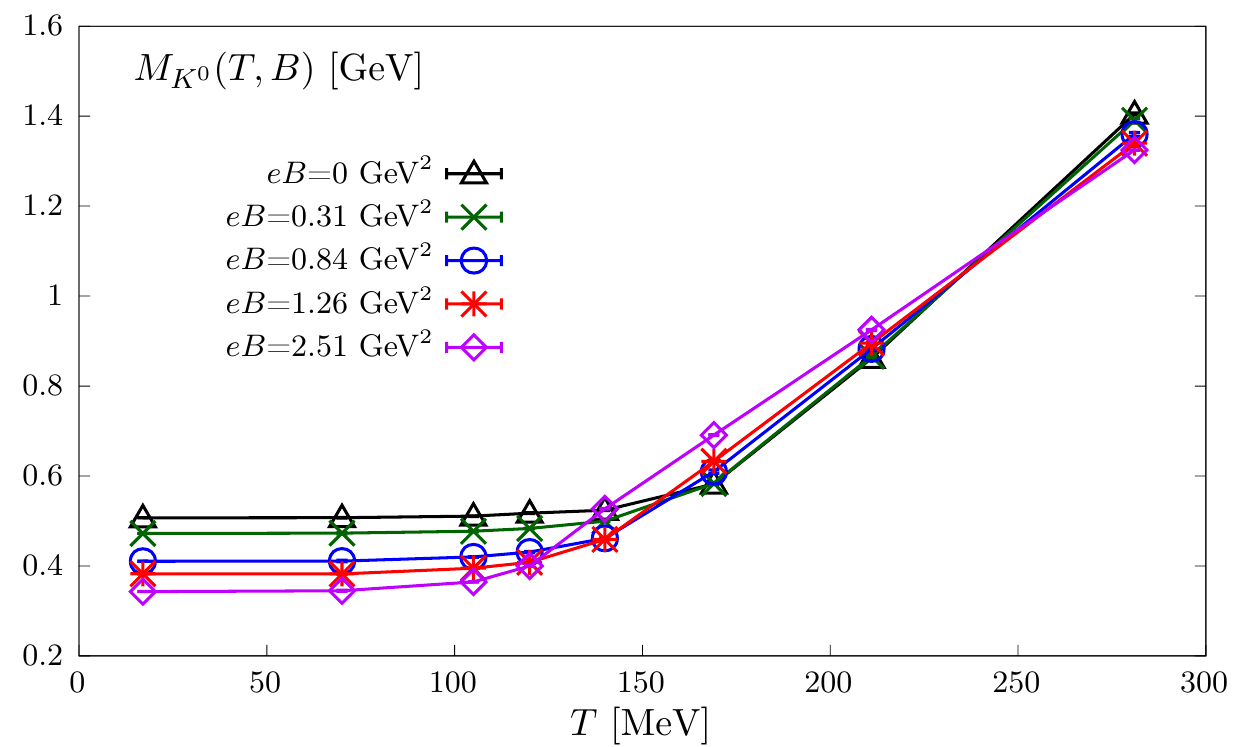}
	\includegraphics[width=0.45\textwidth]{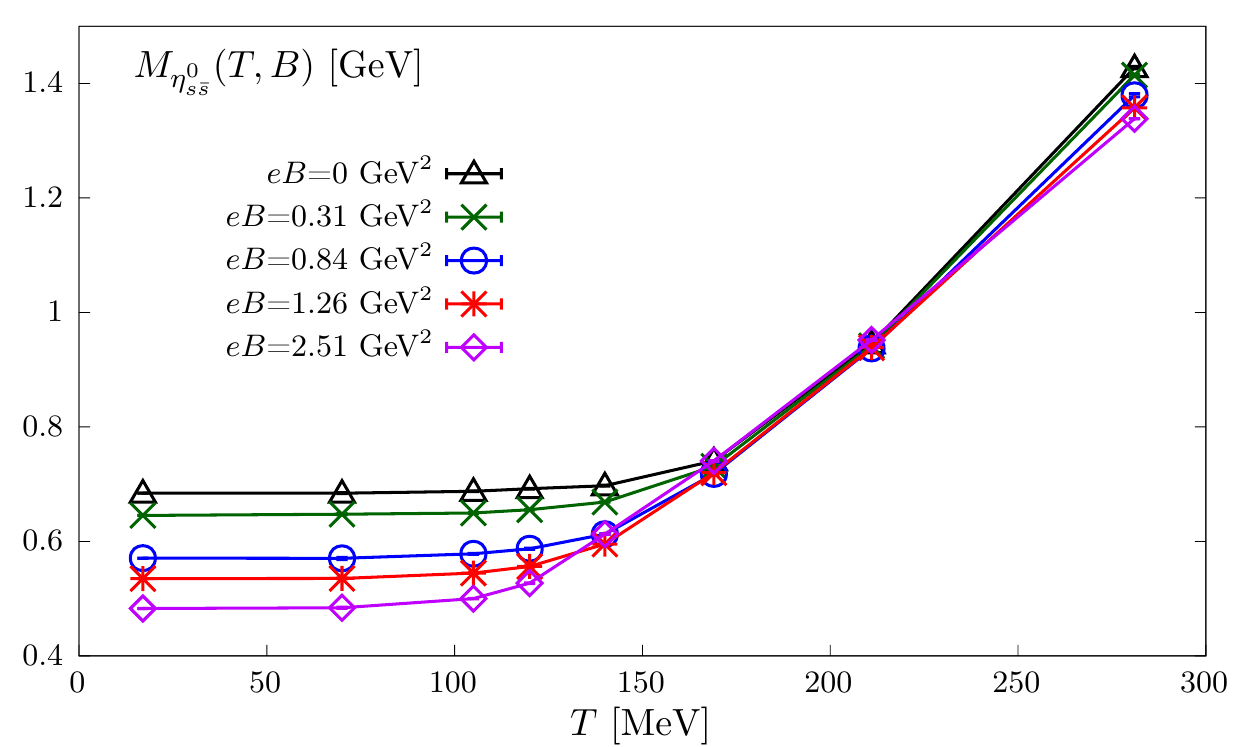}
	\caption{Screening masses of $\pi^0$ (top), $K^0$ (middle) and $\eta^0_{s\bar{s}}$ (bottom) as a function of temperature at several values of $eB$. Straight lines connecting neighboring data points are just used to guide the eye.}
	\label{fig:Mscr-T}
\end{figure}

\begin{figure*}[t]
	\centering
		\includegraphics[width=0.33\textwidth]{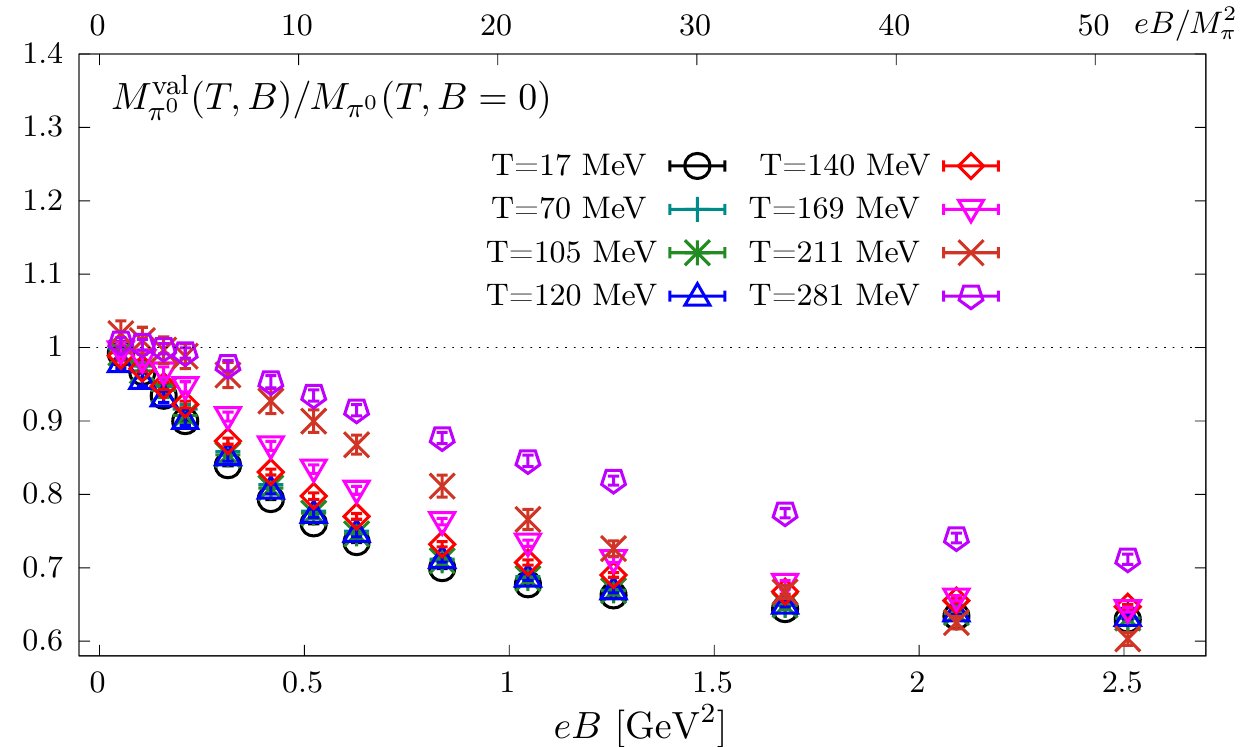}
	\includegraphics[width=0.328\textwidth]{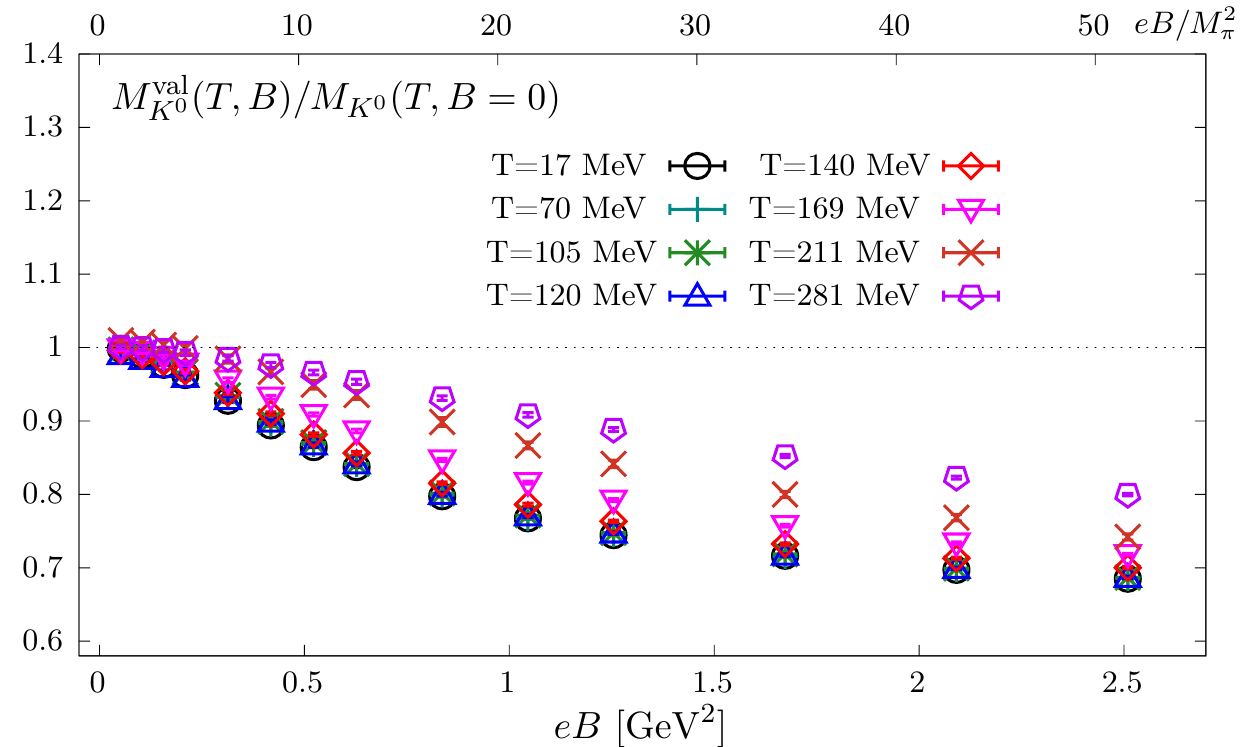}
	\includegraphics[width=0.328\textwidth]{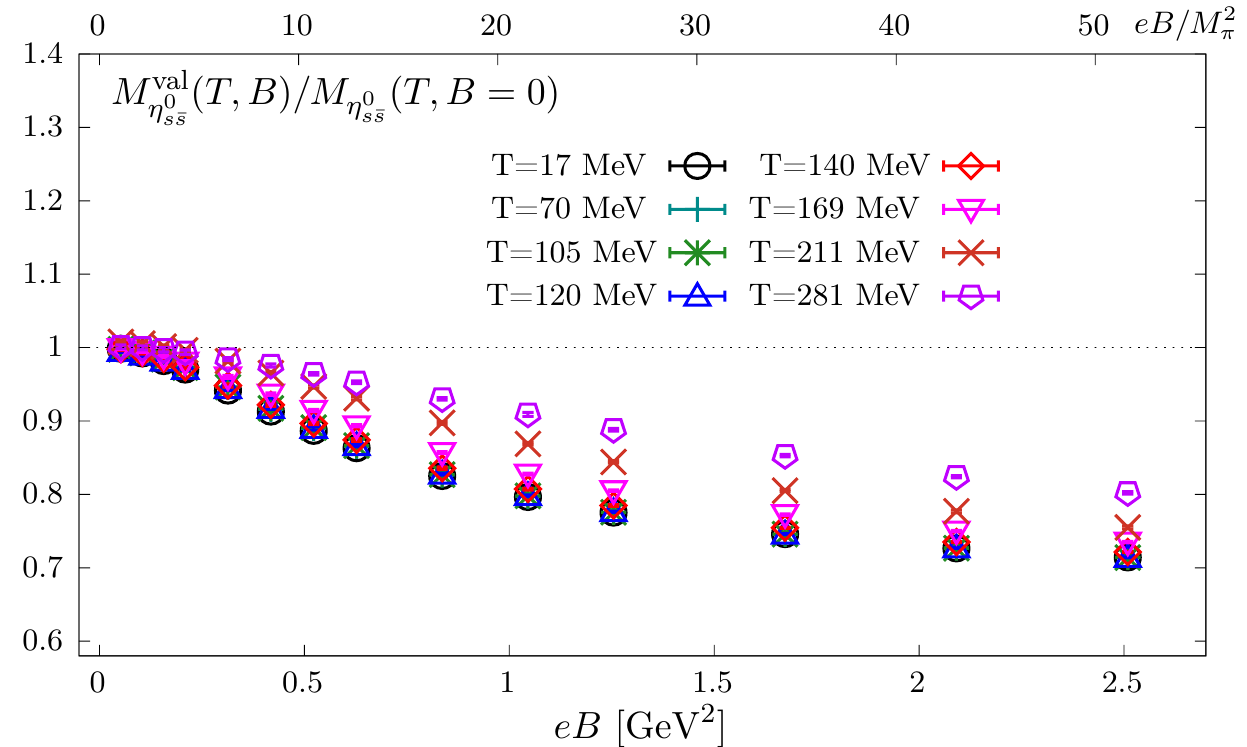} \\
		\includegraphics[width=0.328\textwidth]{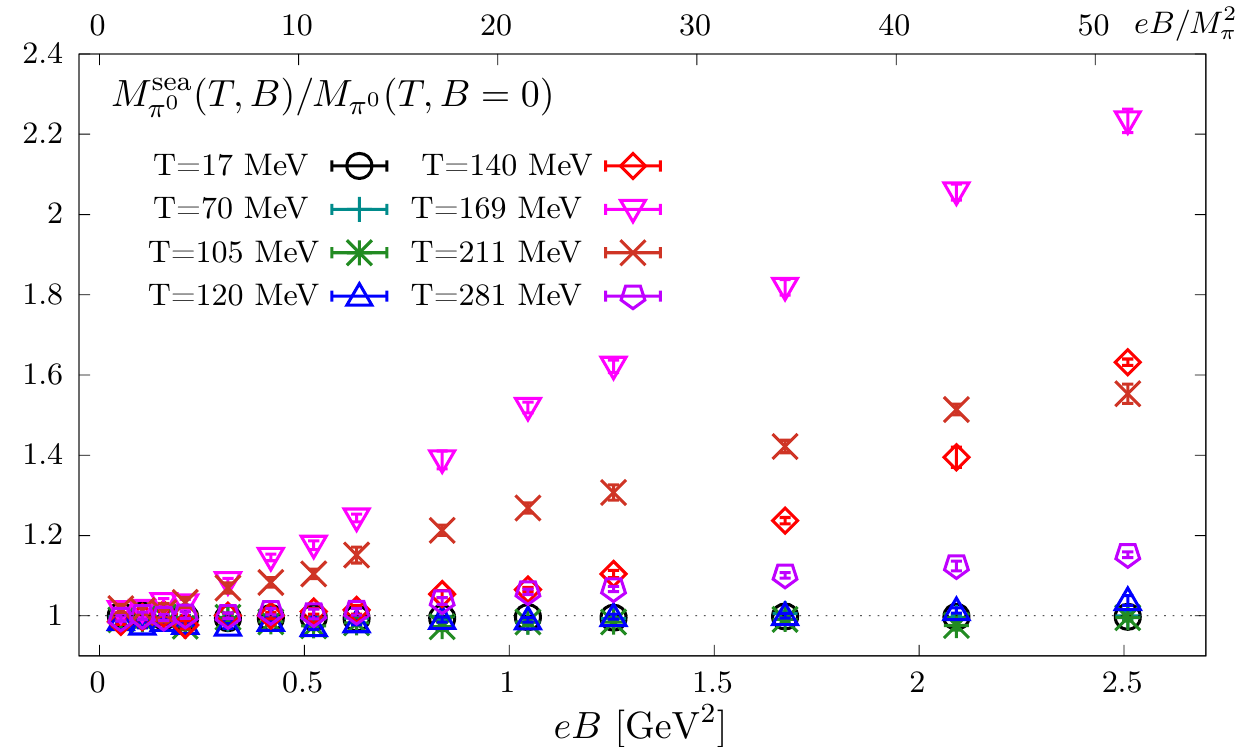}
	\includegraphics[width=0.328\textwidth]{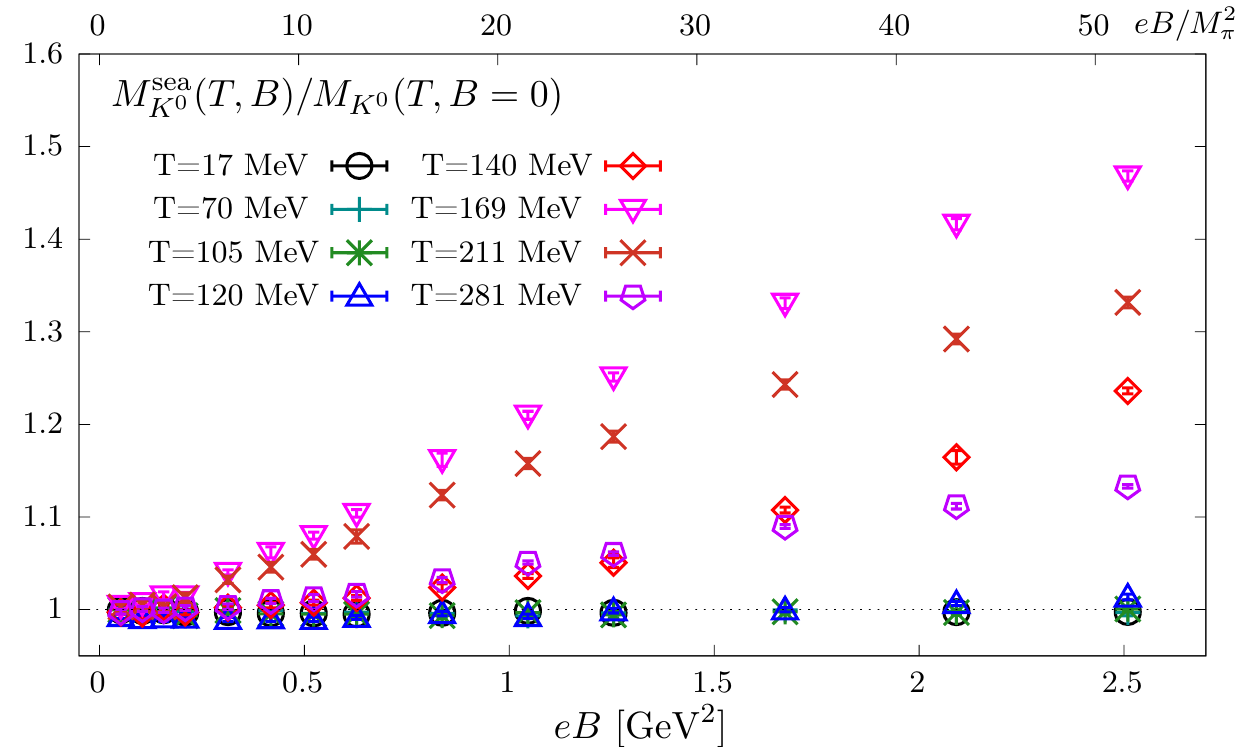}
	\includegraphics[width=0.328\textwidth]{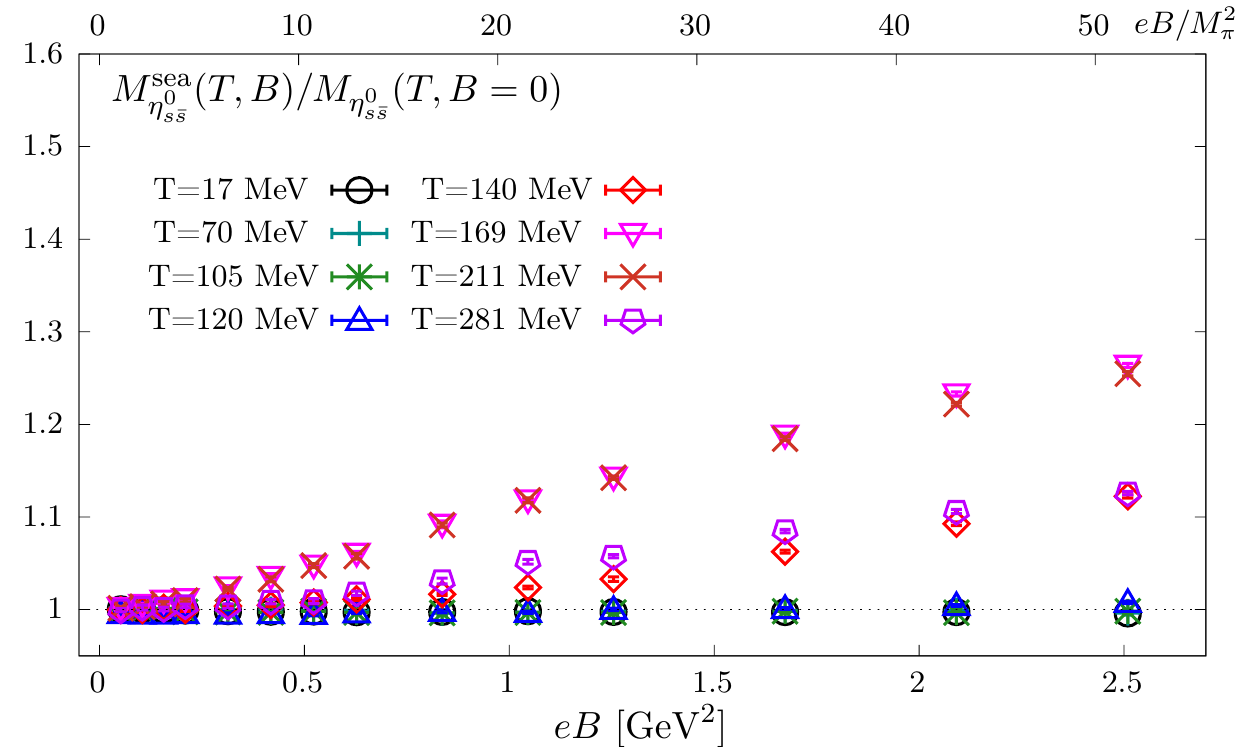}
	\caption{Top: ratios of ``valence" screening masses 
	$M^{\rm val}_H$ for $\pi^0$ (left), $K^0$ (middle) and $\eta^0_{s\bar{s}}$ (right) as a function of $eB$ at various temperatures. Bottom: same as top plots but for the ratios of ``sea" screening masses $M^{\rm sea}_H$.}
	\label{fig:val-sea-eB}
\end{figure*}

\begin{figure*}[!htbp]
	\centering
		\includegraphics[width=0.328\textwidth]{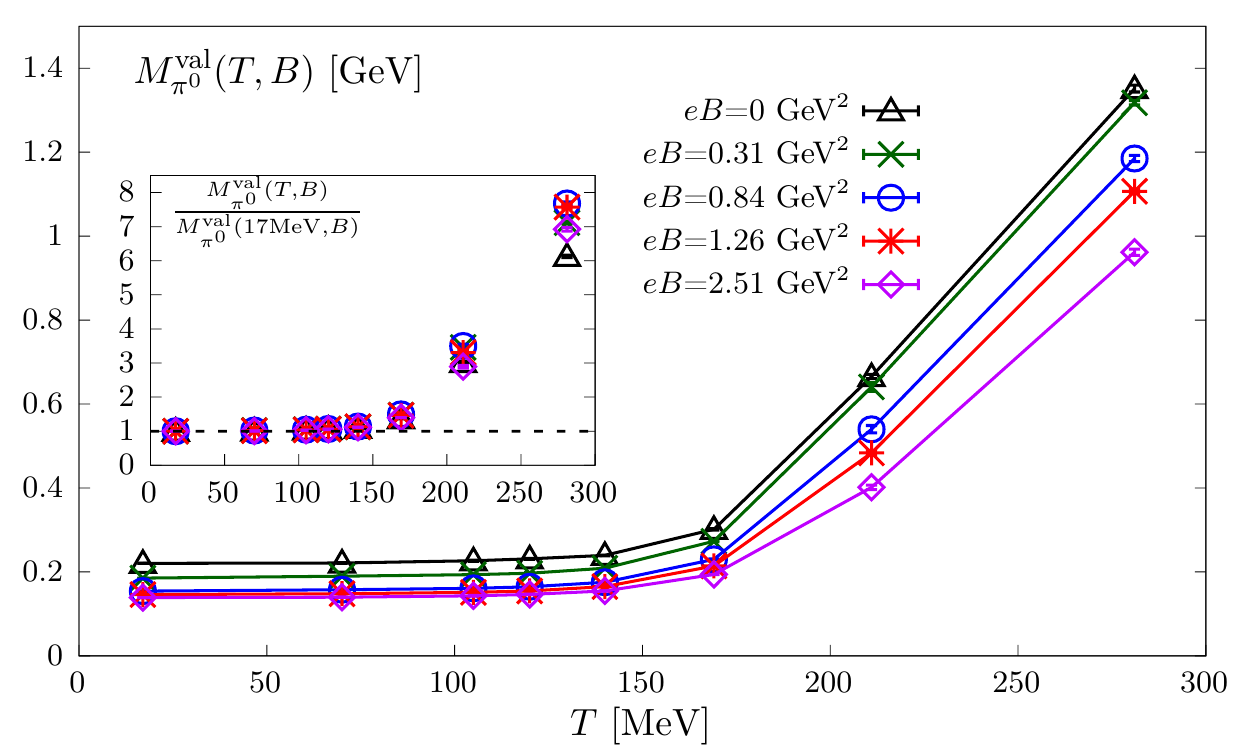}
	\includegraphics[width=0.328\textwidth]{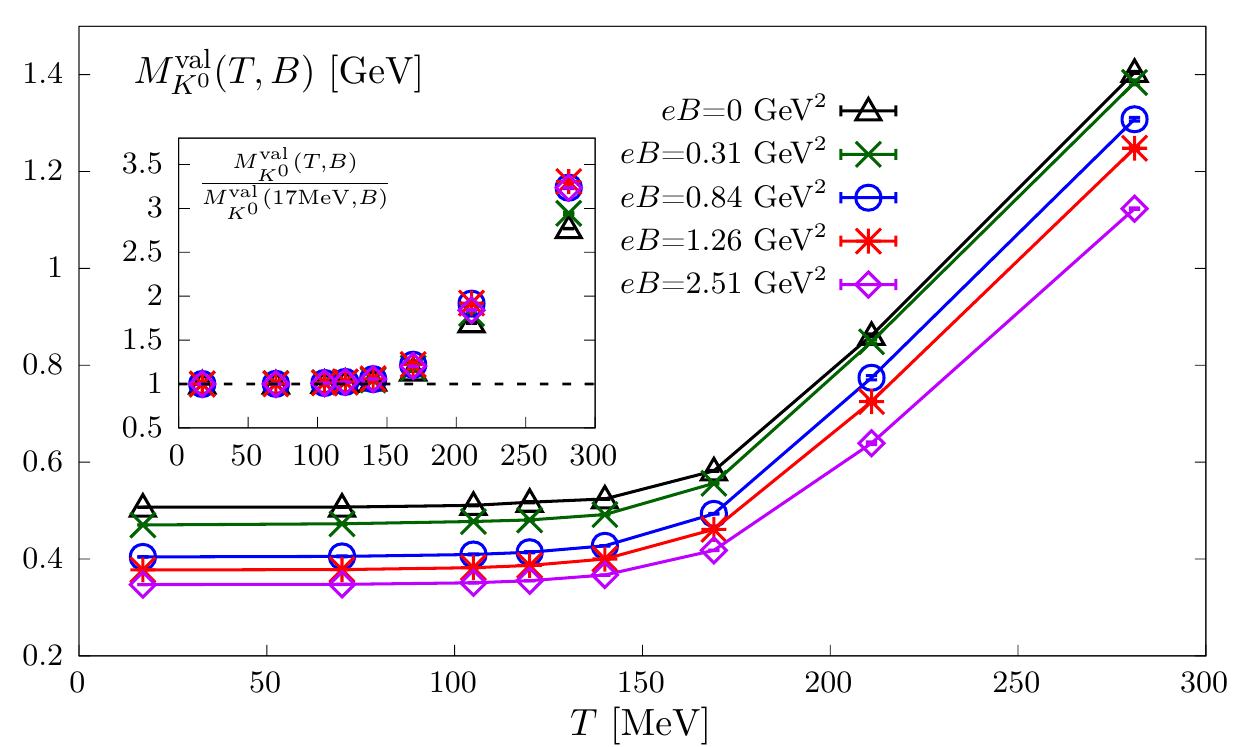}
	\includegraphics[width=0.328\textwidth]{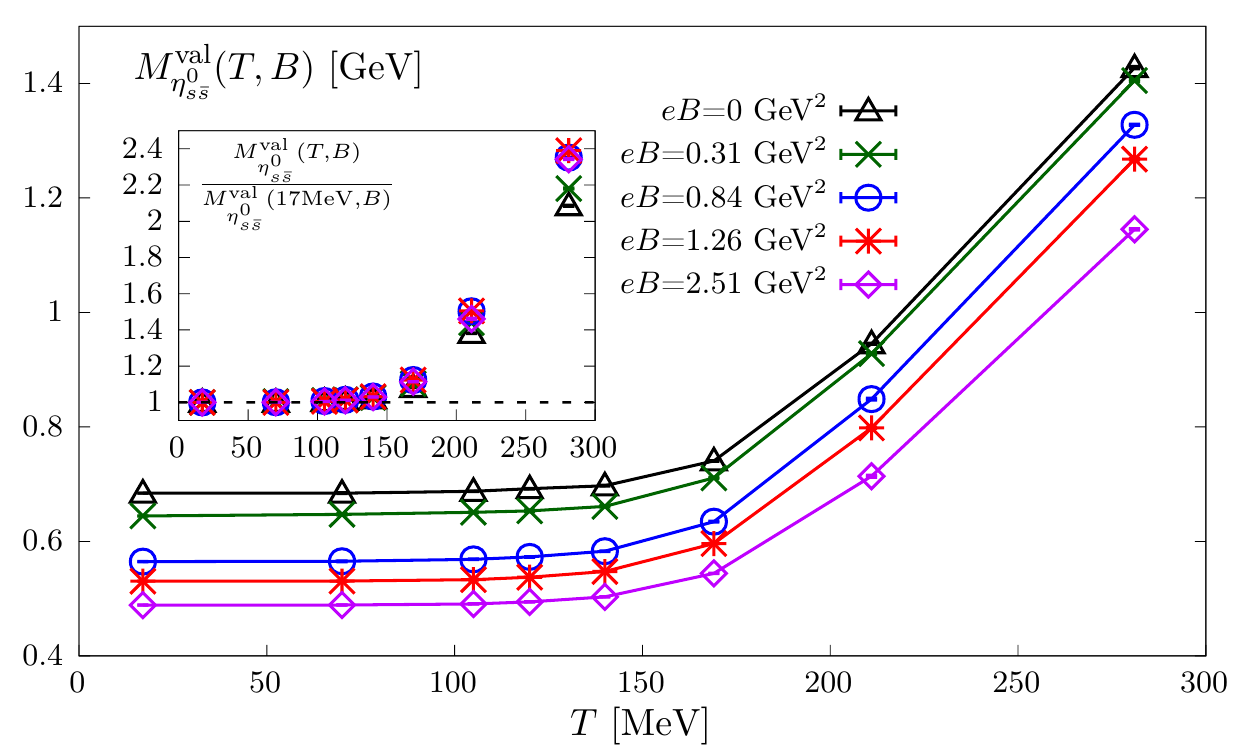} \\
		\includegraphics[width=0.328\textwidth]{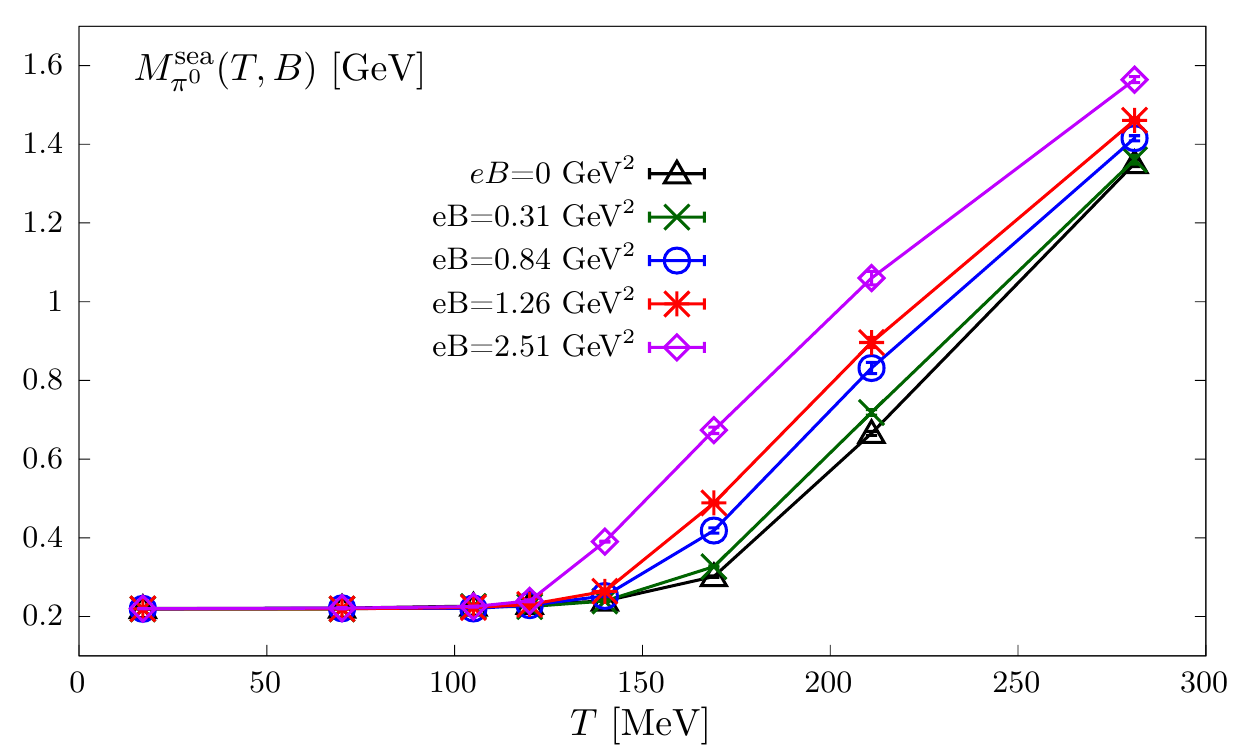}
	\includegraphics[width=0.328\textwidth]{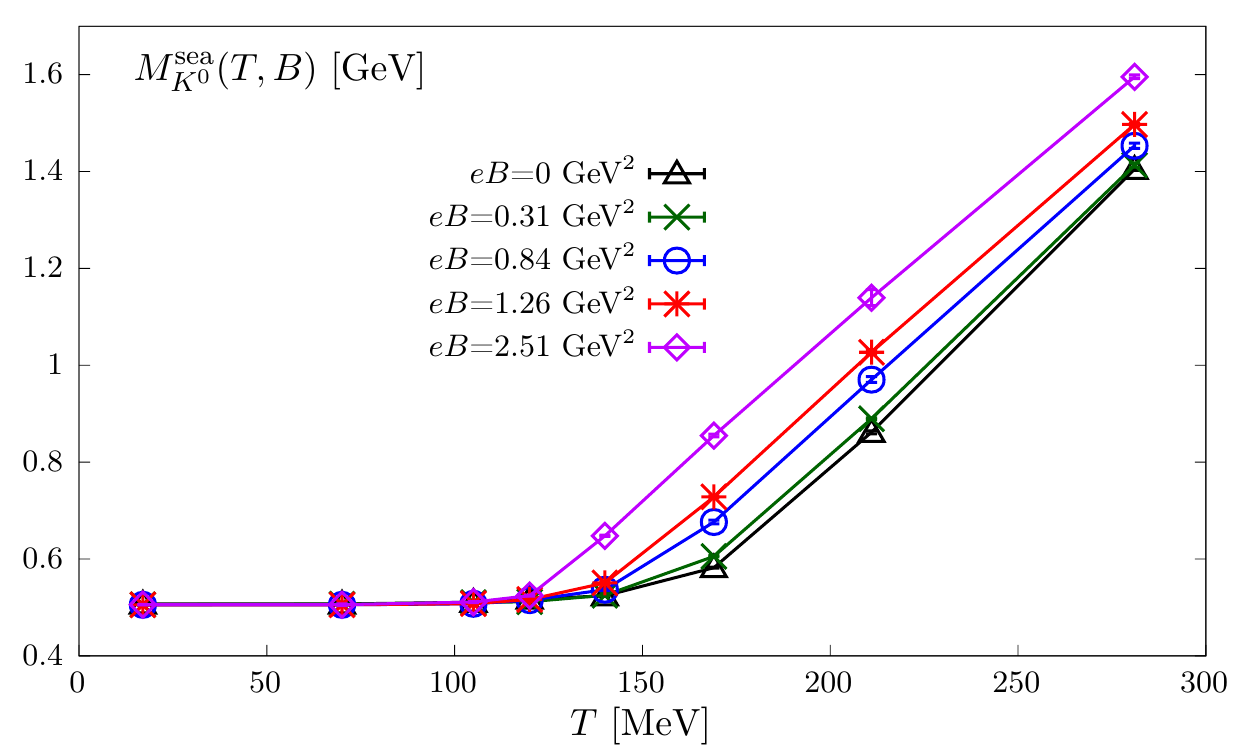}
	\includegraphics[width=0.328\textwidth]{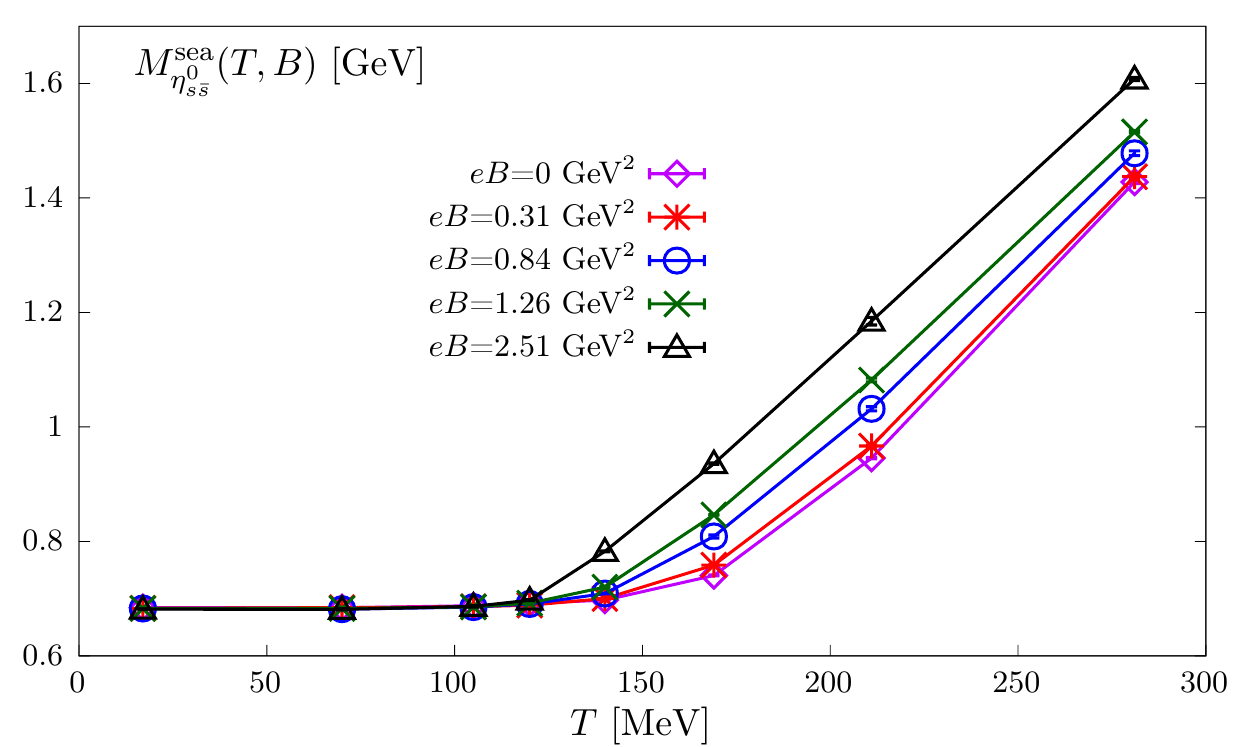}	
	\caption{Top: ``valence" screening masses 
	$M^{\rm val}_H$ for $\pi^0$ (left), $K^0$ (middle) and $\eta^0_{s\bar{s}}$ (right) as a function of $T$ at several values of $eB$. The insets instead show the ratio $M^{\rm val}_H(T,B)/M^{\rm val}_H(T=17~\mathrm{MeV},B)$ as a function of $T$. Bottom: similar to top plots but for ``sea" screening masses $M^{\rm sea}_H$. Straight lines connecting neighboring data points are just used to guide the eye.}
	\label{fig:val-sea-T}
\end{figure*}

In Fig.~\ref{fig:Mscr-eB} (middle) we show a similar plot as Fig.~\ref{fig:Mscr-eB} (top) but for $K^0$. Although the trend of these ratios for $K^0$ in $eB$ at each temperature is similar to those for $\pi^0$, the changes induced by $eB$ are smaller in the screening mass of $K^0$ than that of $\pi^0$. For instance at $T~=~$169 MeV the screening masses are enhanced and the ratios for $K^0$ and $\pi^0$ are about 1.2 and 1.7 at $eB\simeq2.5$ GeV$^2$, respectively, while at $T~=~$17 MeV they are suppressed and the ratios are about 0.68 and 0.6 at $eB\simeq2.5$ GeV$^2$, respectively. This might be due to the fact that $K^0$ is heavier than $\pi^0$, and consequently a same level of change in the screening masses of $K^0$ requires a larger $eB$ compared to that of $\pi^0$.

To further check whether the screening mass of a heavier particle is less affected by $eB$, we show in Fig.~\ref{fig:Mscr-eB} (bottom) the influence of magnetic field to the screening mass of a fictitious neutral pseudoscalar meson $\eta^0_{s\bar{s}}$. In the vacuum at zero magnetic fields $M_{\eta^0_{s\bar{s}}}\simeq 684$ MeV $>$ $M_{K}\simeq507$ MeV $>$ $M_{\pi}\simeq$ 220 MeV in our current lattice setup.  The screening mass of $\eta^0_{s\bar{s}}$ is not enhanced by the magnetic field in the current window of temperature and magnetic field strength, whereas the screening masses of $\pi^0$ and $K^0$ are enhanced by the magnetic field and increase as $eB$ grows in particular at $T$~=~169 and 211 MeV. Note that although the ratio for $\eta^0_{s\bar{s}}$ is not larger than unity, it seems to develop an increasing trend as $eB\gtrsim 1$ GeV$^2$ at $T$~=~140 and 169 MeV.

In Fig.~\ref{fig:Mscr-T} we show the temperature dependence of screening masses of $\pi^0$ (top), $K^0$ (middle) and $\eta^0_{s\bar{s}}$ (bottom) at several fixed values of $eB$. At $eB=0$ all the screening masses 
remain almost independent of temperature at $T\lesssim140$ MeV $\sim 0.8T_{pc}(eB=0)$. They then suddenly jump to a larger value at $T$~=~169 MeV$\sim T_{pc}(eB=0)$ and increases as $T$ grows. This is compatible with studies in lattice QCD with physical pion mass~\cite{Bazavov:2019www}. As the magnetic field is turned on all the screening masses show a trend that they jump at a lower temperature with a larger $eB$. This suggests the reduction of $T_{pc}$ in stronger magnetic fields, which is consistent with the observation from the inflection points of chiral condensates shown in Fig.~\ref{fig:PBP-T}. Note again that although the screening mass of $\eta^0_{s\bar{s}}$ is not enhanced by the magnetic field its temperature dependence still suggests the reduction of $T_{pc}$ in stronger magnetic fields. It is also interesting to point out that 
the vacuum mass of $\eta^0_{s\bar{s}}$, $M_{\eta^0_{s\bar{s}}}\simeq 684$ MeV, in our current lattice setup is larger than $520$ MeV. With the value of pion mass larger than 520 MeV it is found that QCD turns from displaying inverse magnetic catalysis to magnetic catalysis of light quark condensates while $T_{pc}$ still decreases as $eB$ grows~\cite{DElia:2018xwo,Endrodi:2019zrl}.

 In Fig.~\ref{fig:val-sea-eB} we show the valence and sea quark effects to the ratios shown in Fig.~\ref{fig:Mscr-eB}. The $eB$ dependences of ratios of the screening mass of $\pi^0$ (left), $K^0$ (middle) and $\eta^0_{s\bar{s}}$ to their corresponding values at $eB=0$ are obtained from ${G}^{\rm val}_H(B,T,z)$ [top, cf. Eq.~\eqref{eq:val-corr}] and ${G}^{\rm sea}_H(B,T,z)$ [bottom, cf. Eq.~\eqref{eq:sea-corr}], respectively. For the valence quark effects it can be clearly seen that the ratio decreases as $eB$ grows at each temperature, while the ratio from the sea quark effects increases as $eB$ grows at $T\gtrsim 120$ MeV and remains as unity at $T<$ 120 MeV. Again both the sea and valence quark effects are less significant in heavier mesons. It can also be observed that the sea quark effects are more sensitive to the quark mass compared to the valence quark effects. Thus these two effects compete with each other and lead to the results shown in Fig.~\ref{fig:Mscr-eB}.

In Fig.~\ref{fig:val-sea-T} we further show screening masses of $M^{\rm val}_H$ (top) and $M^{\rm sea}_H$ (bottom) for $\pi^0$ (left), $K^0$ (middle) and $\eta^0_{s\bar{s}}$ (right) as a function of temperature at several fixed values of $eB$. It can also be clearly seen that $M^{\rm val}_H$ are suppressed at all temperatures while $M^{\rm sea}_H$ is enhanced at high temperatures and remains almost unaffected at low temperatures by the magnetic field. As observed from the insets in Fig.~\ref{fig:val-sea-T} (top), the temperature where $M^{\rm val}_H$ increases most rapidly seems to be independent of $eB$. On the other hand, it is obvious that the temperature where $M^{\rm sea}_H$ increases most rapidly decreases with increasing $eB$. In other words, the reduction of $T_{pc}$ in stronger magnetic fields is only manifested in the temperature dependence of $M^{\rm sea}_H$. This, however, is also obvious in the sense that the transition temperature is one of the thermodynamic properties that is encoded in the partition function and manifested in the ``sea quark" relevant quantities. The same conclusions can be drawn from the sea and valence quark effects to chiral condensates shown in Appendix~\ref{sec:app}.

\section{Conclusion and discussion}
\label{sec:summary}
In this work we pointed out that the $eB$ and temperature dependences of chiral condensates are intrinsically connected to screening masses of the neutral pseudoscalar mesons. We have demonstrated this, to the best of our knowledge, for the first time via the first principle lattice QCD simulations. The observed complex dependences of chiral condensates on $eB$ and $T$ actually reflect the change of screening length (mass) of corresponding neutral pseudoscalar mesons. These complex dependences are attributed to the competition between sea and valence quark effects. The former effect tends to enhance the screening mass while the latter one tends to suppress the screening mass.

We find that the influence of $eB$ becomes smaller to the heavier neutral meson and associated quark chiral condensates. As the neutral meson is sufficiently heavy, the inverse magnetic catalysis of corresponding quark chiral condensates ceases to occur. This is the case of $\eta^0_{s\bar{s}}$ [$M_{\eta^0_{s\bar{s}}}(T=0,eB=0)\simeq~684$ MeV] and the corresponding strange quark chiral condensate. On the other hand, the reduction of $T_{pc}$ in the magnetic field always holds based on the temperature dependences of up, down and strange quark chiral condensates as well as screening masses of $\pi^0$, $K^0$ and $\eta^0_{s\bar{s}}$. This reflects the crossover nature of the QCD transition in the current temperature and magnetic field.

The reduction of $T_{pc}$ accompanying with the absence of inverse magnetic catalysis of light quark chiral condensates was also found in previous lattice QCD studies with the pion masses $500 \lesssim M_\pi(eB=0)\lesssim 660$ MeV using stout improved staggered fermions on $N_\tau=6$ lattices~\cite{DElia:2018xwo,Endrodi:2019zrl}. This is consistent with our current findings and could be explained as follows. The transition temperature is  manifested in the sea quark relevant quantities, and its reduction in the strong magnetic field seems to hold true as long as the sea quark effects exist. However, the display of inverse and magnetic catalyses in QCD is more due to a competition between the sea and valence quark effects. If sea quark effects win the inverse magnetic catalysis of quark chiral condensates occurs, while if valence quark effects win the magnetic catalysis occurs. In the heavy quark mass limit, i.e. in the case of quenched QCD and there exist no sea quark effects, the QCD transition becomes first order and the phase transition temperature $T_c$ will become independent of $eB$ as gluons are blind to $eB$. Meanwhile, in the quenched QCD there are no ``sea" quark chiral condensates, and the ``valence" quark chiral condensates are expected to always get catalyzed by the magnetic field and lead to a constant $T_{c}$.

Our simulations of $N_f=2+1$ QCD are performed using highly improved staggered fermions with larger-than-physical pion mass, i.e. $M_\pi=220$ MeV at $eB=0$ and with a single lattice spacing $a\simeq$ 0.117 fm. As the $eB$ and temperature dependence of our results of light quark chiral condensates are compatible with continuum extrapolated results obtained using physical pion masses and $a\in[0.1,0.29]$ fm~\cite{Bali:2012zg}, the lattice cutoff effects in our study should be small. Nevertheless, further studies on the current complex behavior of screening masses using lattice QCD with physical pion masses in the continuum limit would be important in the future. On the other hand, in the very strong magnetic field the transition of QCD at the physical point is expected to become first order~\cite{Endrodi:2015oba}, it would be interesting to study the screening mass in such a strong magnetic field where the first order phase transition was observed very recently~\cite{DElia:2021yvk}.

%------------------------------------------------------------------------------------
%  acknowledgments
%------------------------------------------------------------------------------------
\section*{Acknowledgements}
We thank Defu Hou, Toru Kojo and Swagato Mukherjee for interesting discussions. This work was supported by the National Natural Science Foundation of China under Grant No. 11775096, and the Guangdong Major Project of Basic and Applied Basic Research No.
2020B0301030008. The numerical simulations have been performed on the Graphics Processing Unit (GPU) cluster in the Nuclear Science Computing Center at Central China Normal University (NSC$^3$), Wuhan, China.

\begin{widetext}
\appendix

\section{Valence and sea quark effects to up, down and strange quark chiral condensates}
\label{sec:app}

In Fig.~\ref{fig:PBP_val-sea} we show valence (top) and sea (bottom) quark effects to the change of up, down and strange quark chiral condensates as a function of $eB$ at various $T$. Figure~\ref{fig:PBP_val-sea-T} is similar to Fig.~\ref{fig:PBP_val-sea} but is plotted as a function of $T$ at several values of $eB$. $\Delta\Sigma^{\rm val}_{ud,ds,s}$ and $\Delta
\Sigma^{\rm sea}_{ud,ds,s}$ are defined in the same spirit as those for correlation functions [cf. Eq.~\eqref{eq:val-corr} and Eq.~\eqref{eq:sea-corr}]. Results of $\Delta\Sigma^{\rm val,sea}_{ud,ds,s}$ shown in Figs.~\ref{fig:PBP_val-sea} and~\ref{fig:PBP_val-sea-T} are actually obtained from $\chi^{\rm val, sea}_H$ using the Ward-Takahashi identities. As seen from Figs.~\ref{fig:PBP_val-sea} and ~\ref{fig:PBP_val-sea-T}, the same conclusions on the role of valence and sea quark effects to the (inverse) magnetic catalysis and reduction of $T_{pc}$ can be drawn as those drawn from screening masses.

\begin{figure*}[!htbp]
	\centering
		\includegraphics[width=0.328\textwidth]{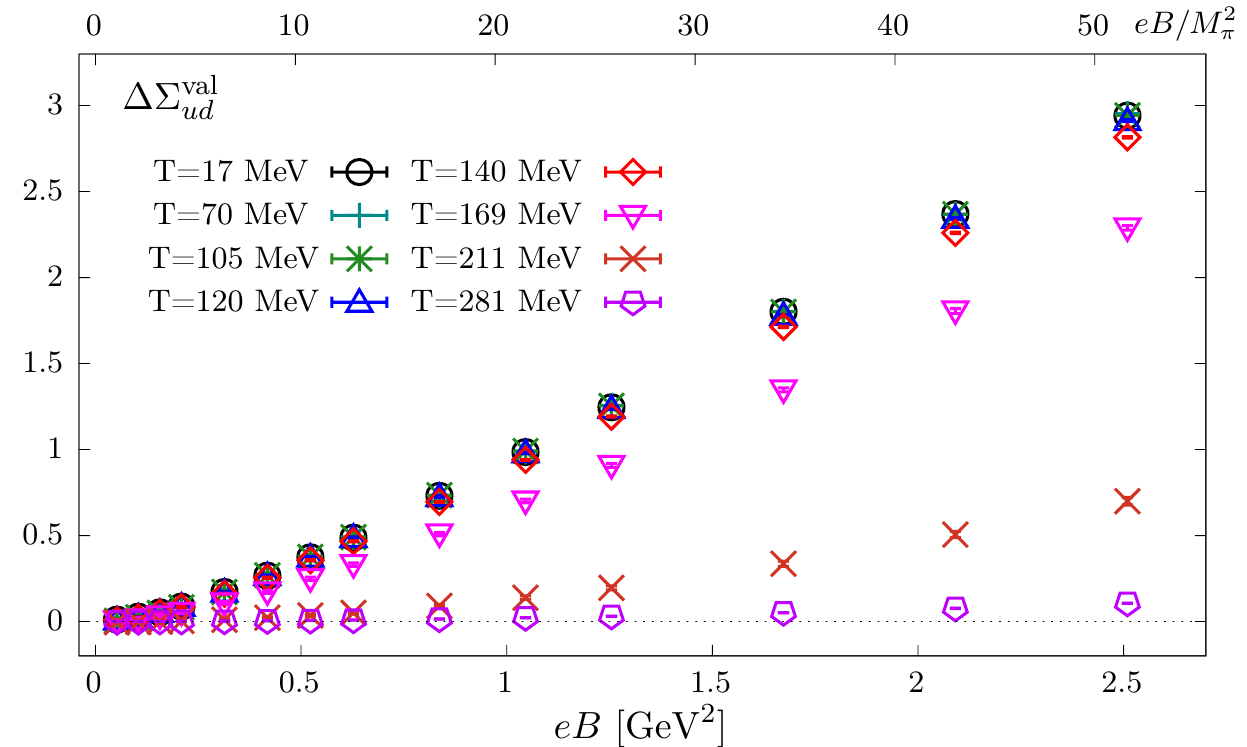}
	\includegraphics[width=0.328\textwidth]{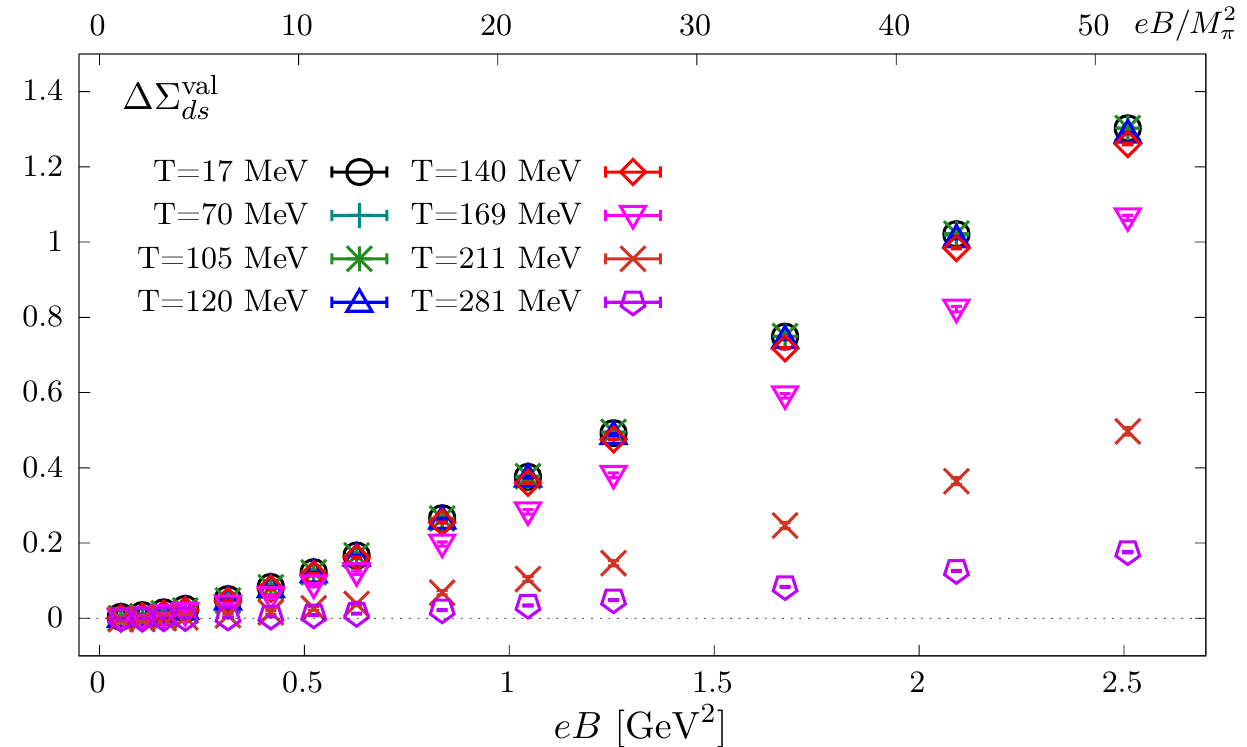}
	\includegraphics[width=0.328\textwidth]{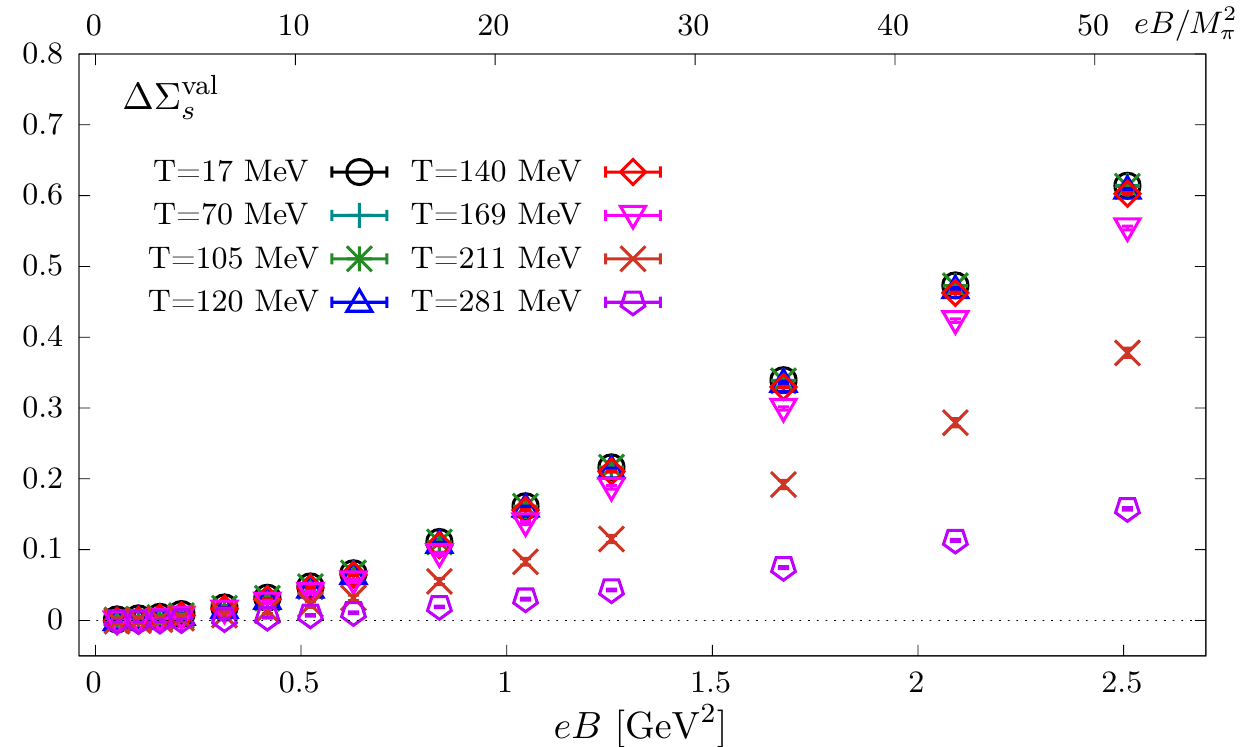} \\
		\includegraphics[width=0.328\textwidth]{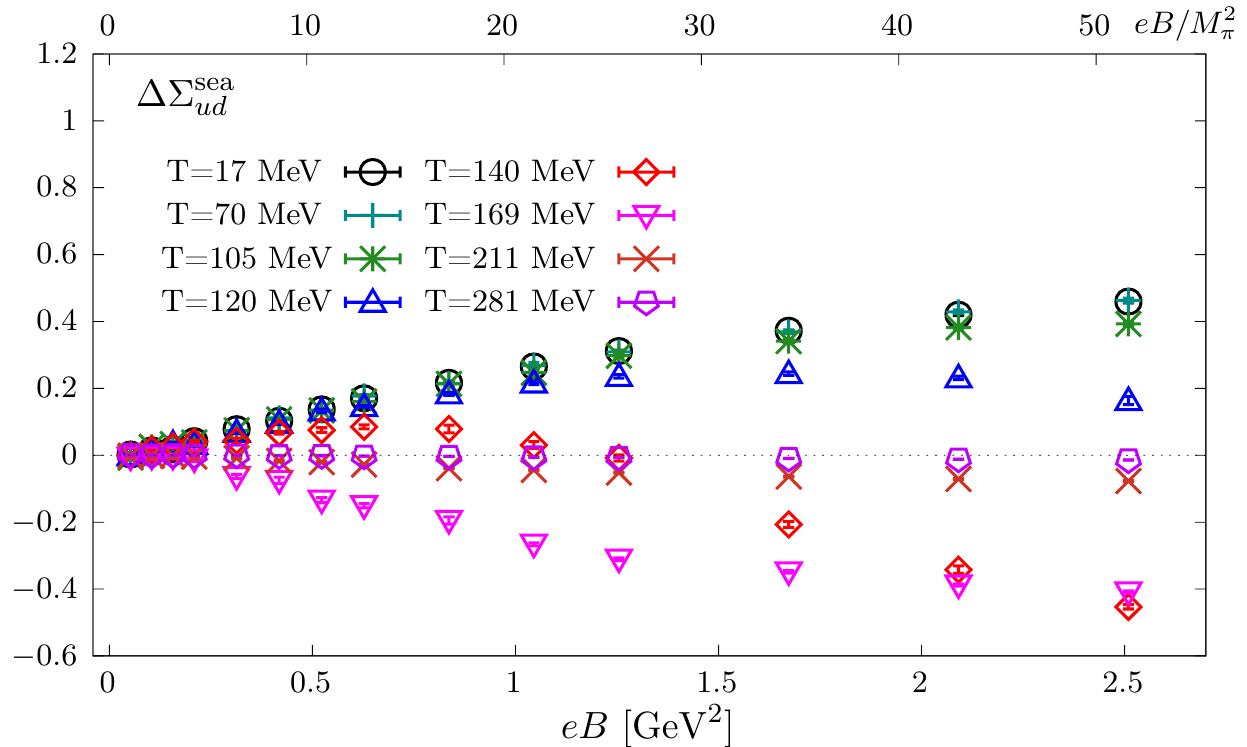}
	\includegraphics[width=0.328\textwidth]{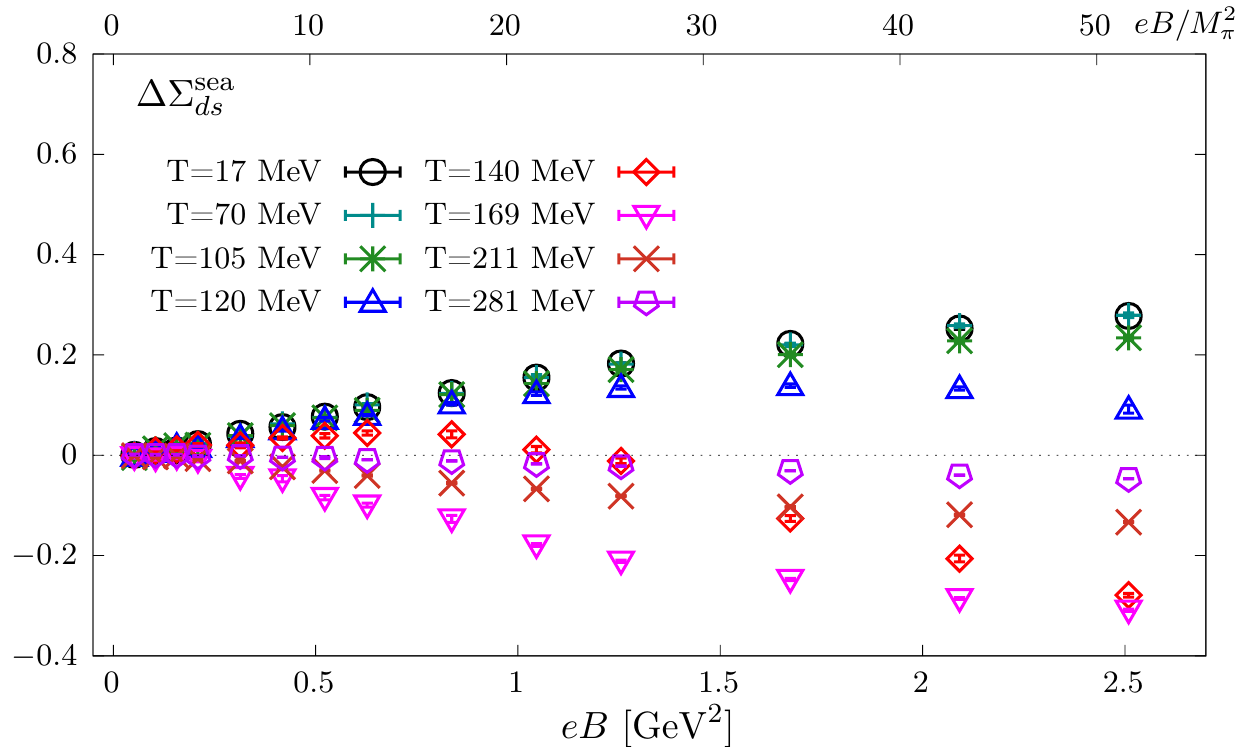}
	\includegraphics[width=0.328\textwidth]{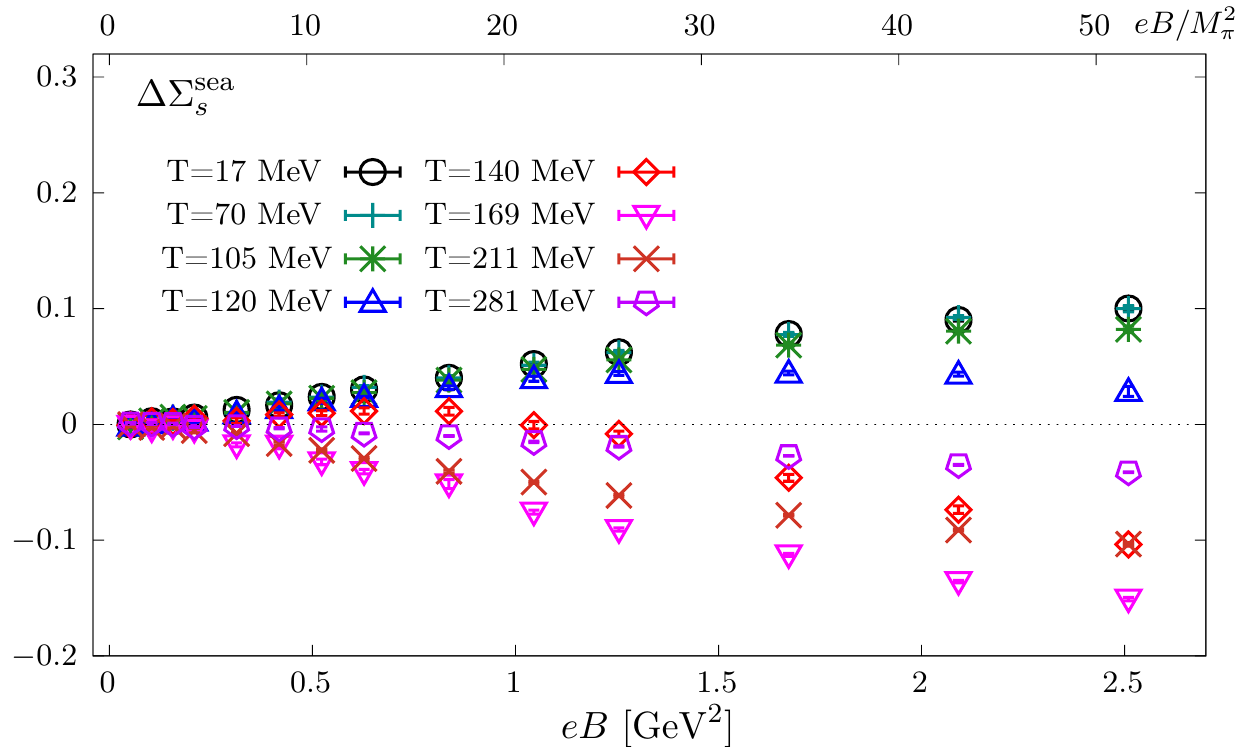} 
	\caption{Top: changes of ``valence" quark chiral condensates,
	$\Delta\Sigma_{ud}^{\rm val}$ (left), $\Delta\Sigma_{ds}^{\rm val}$  (middle) and $\Delta\Sigma_{s}^{\rm val}$  (right) as a function of $eB$ at various values of $T$. Bottom: similar to top plots but for those of ``sea" quark chiral condensates.}
	\label{fig:PBP_val-sea}
\end{figure*}
\begin{figure*}[!htbp]
	\centering
			\includegraphics[width=0.328\textwidth]{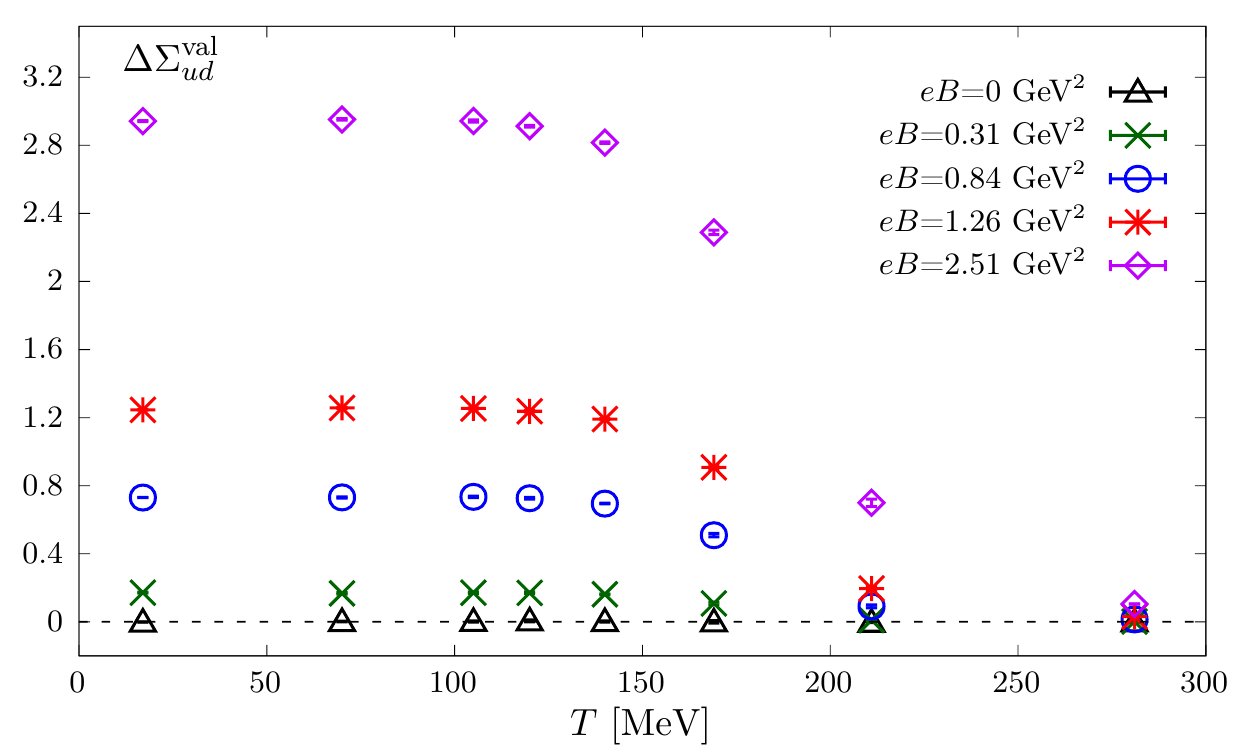}
	\includegraphics[width=0.328\textwidth]{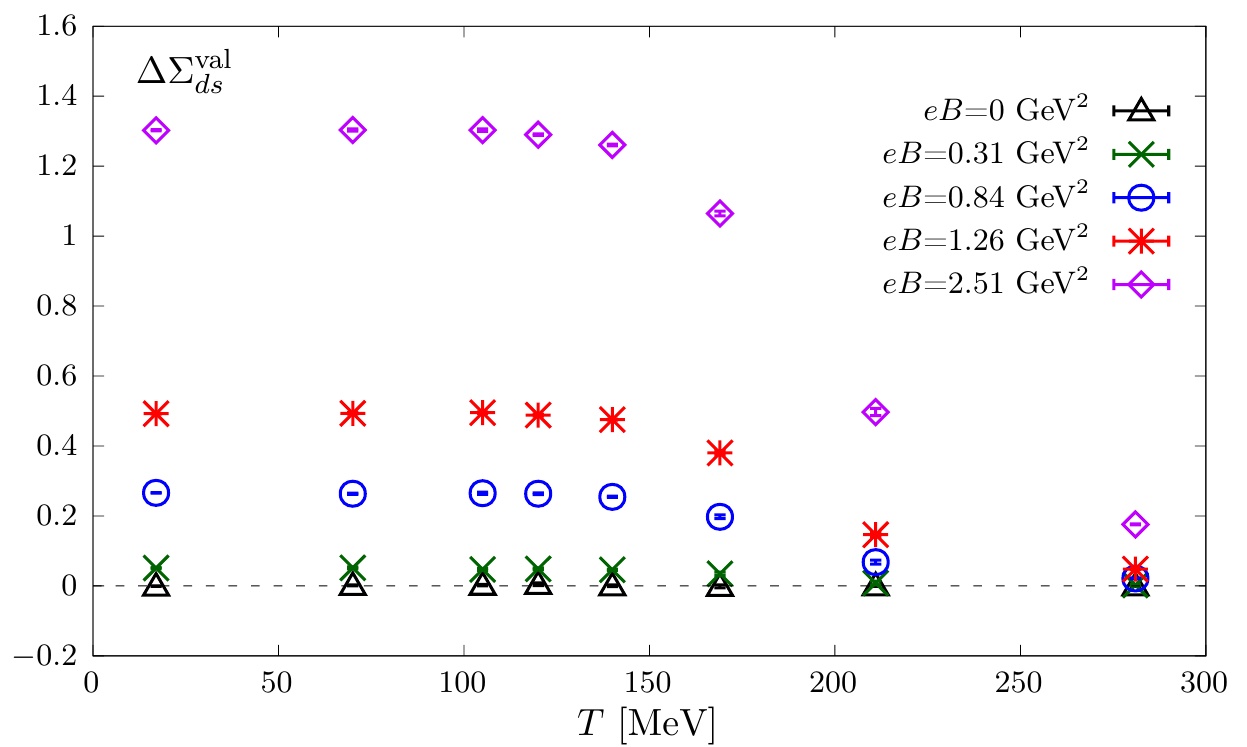}
	\includegraphics[width=0.328\textwidth]{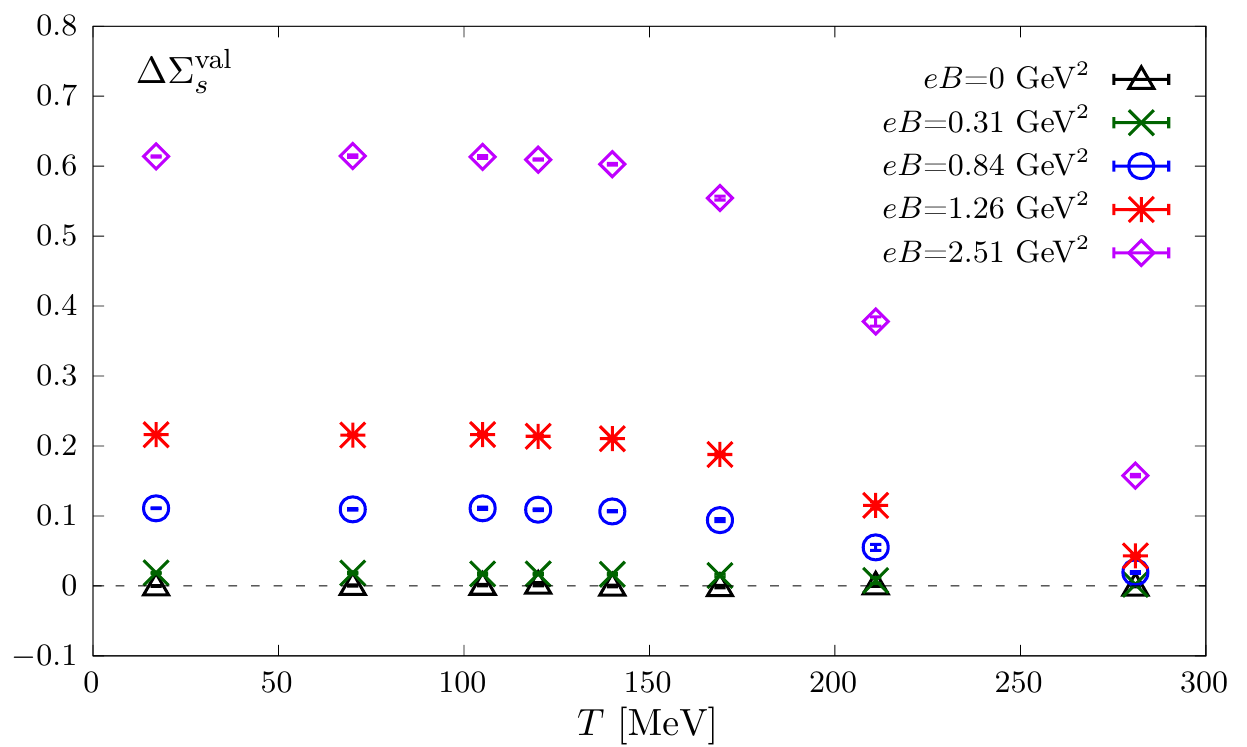} \\
		\includegraphics[width=0.328\textwidth]{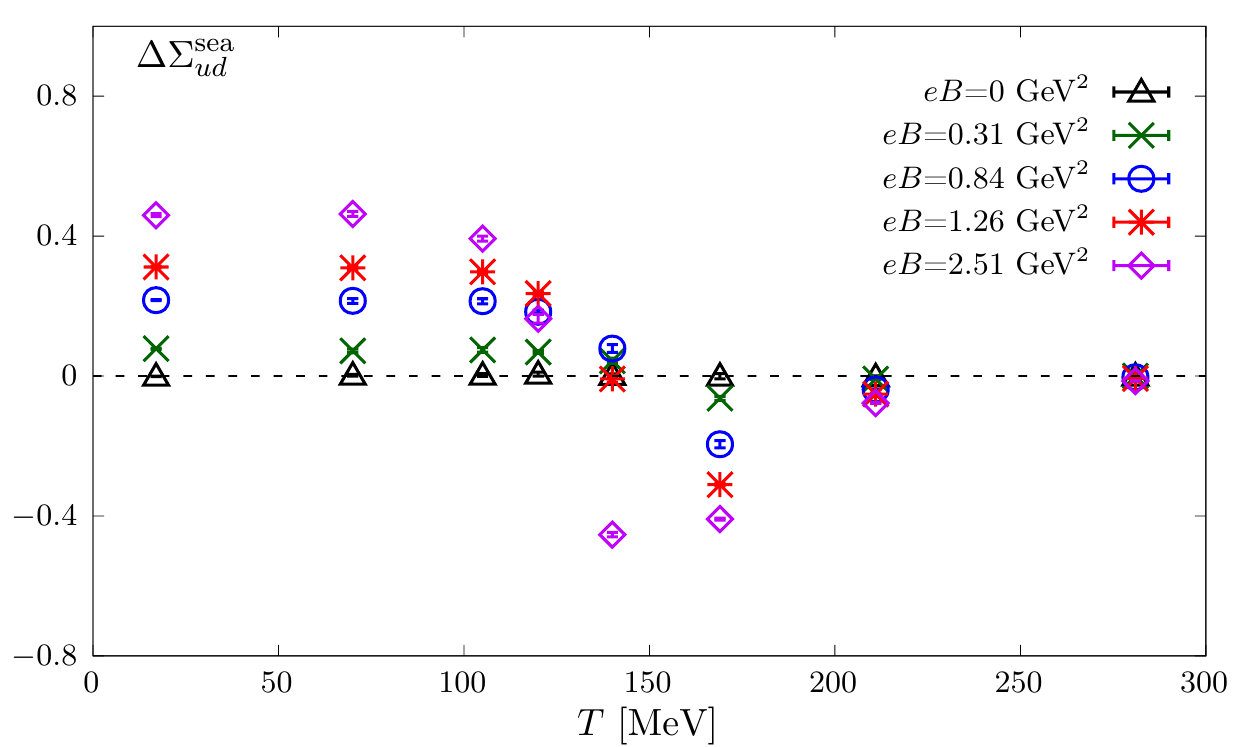}
	\includegraphics[width=0.328\textwidth]{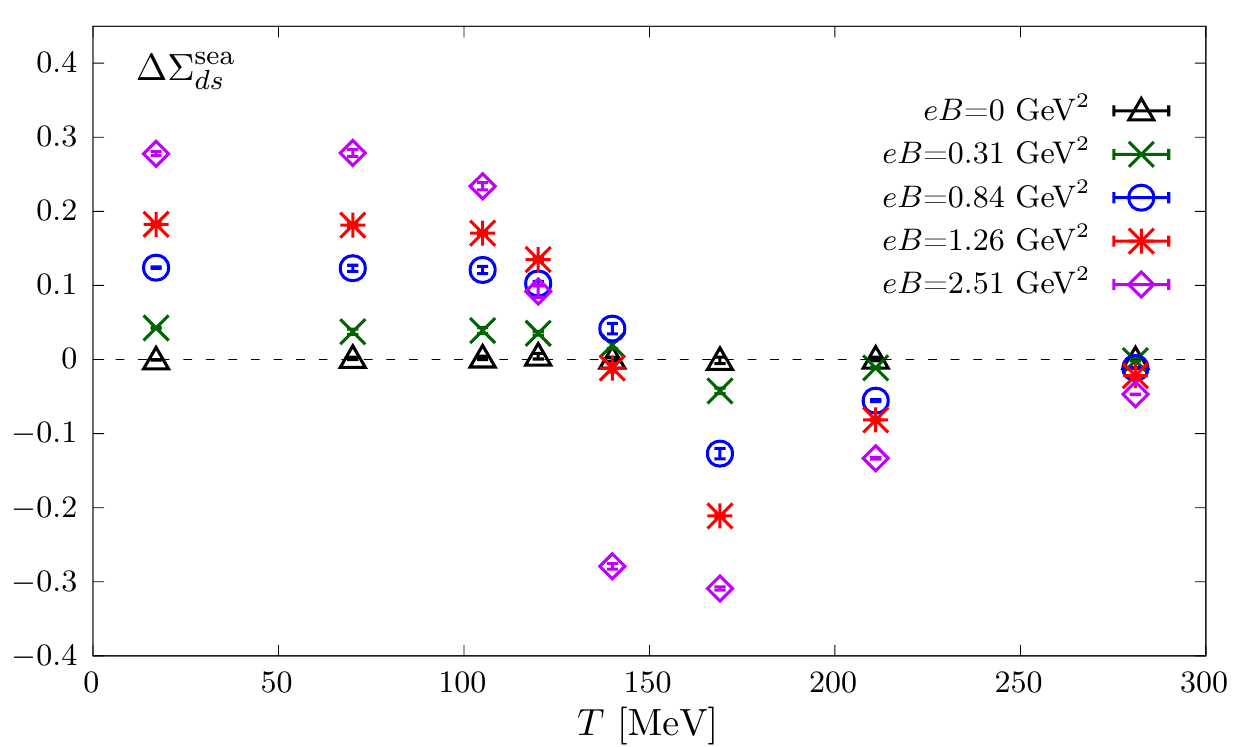}
	\includegraphics[width=0.328\textwidth]{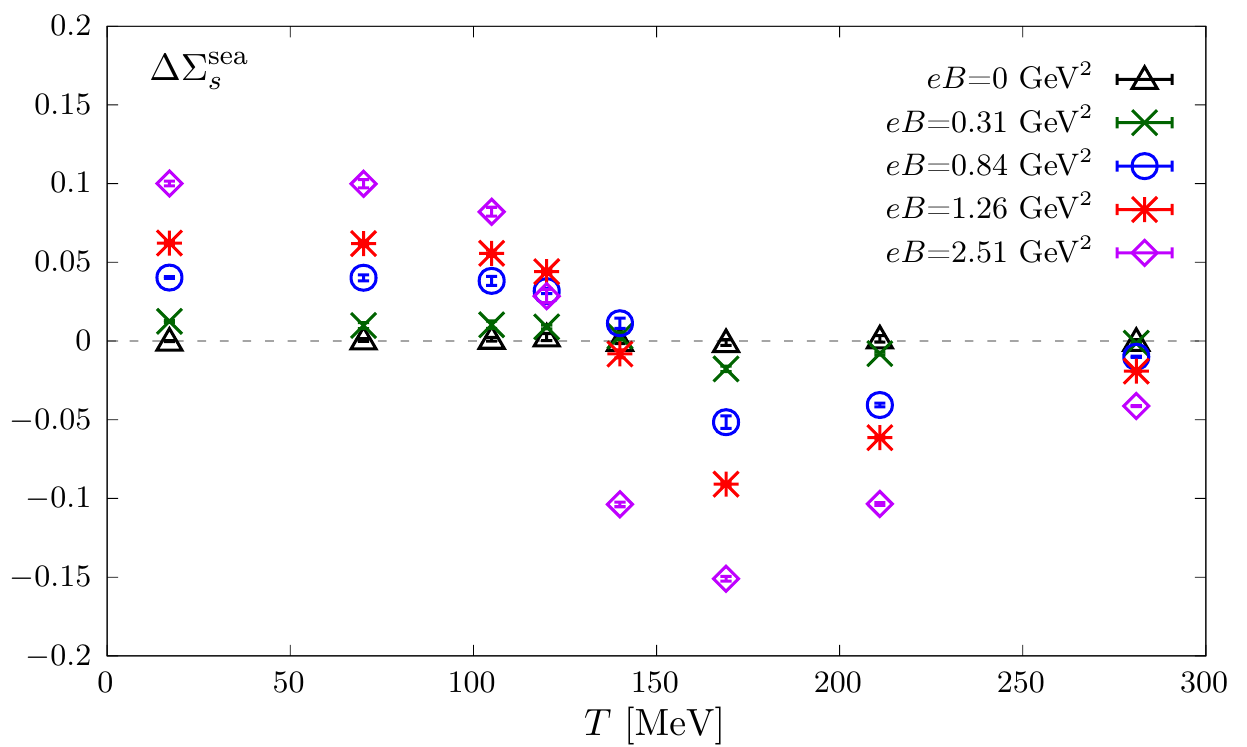} 
	\caption{Top: changes of ``valence" quark chiral condensates,
	$\Delta\Sigma_{ud}^{\rm val}$ (left), $\Delta\Sigma_{ds}^{\rm val}$ (middle) and $\Delta\Sigma_{s}^{\rm val}$  (right) as a function of $T$ at several values of $eB$. Bottom: similar to top plots but for those of ``sea" quark chiral condensates.}
	\label{fig:PBP_val-sea-T}
\end{figure*}

\section{Extraction of screening masses from spatial correlation functions}
\label{sec:app_method}
The method for the extraction of screening masses has been described in detail in Refs.~\cite{Bazavov:2019www,Ding:2020hxw}. Here we show the extraction at $T$~=~140 MeV as a typical example in Fig.~\ref{fig:app_plateau}.
The following ansatz is adopted to fit the spatial correlation function~\cite{Bazavov:2019www}:
\begin{align}
G_H\left(n_z\right)=\sum_{i}\Bigg [ A_{H, i} \cosh\left(aM_{H,i}\left(n_z-\frac{N_s}{2}\right)\right) - (-1)^{n_{z}}  \tilde{A}_{H, i} \cosh\left(a\tilde{M}_{H,i}\left(n_z-\frac{N_s}{2}\right)\right) \Bigg]  \,,
\end{align}
where $n_z=z/a$. 
We choose the plateau of screening mass (the lowest value of $M_{H,i}$) in the pseudoscalar channel based on the Akaike information criterion (AICc), and then obtain the final screening mass and its uncertainty from the plateau using a Gaussian bootstrapping method~\cite{Bazavov:2019www,Ding:2020hxw}.

\begin{figure*}[!htbp]
\centering
	\includegraphics[width=0.328\textwidth]{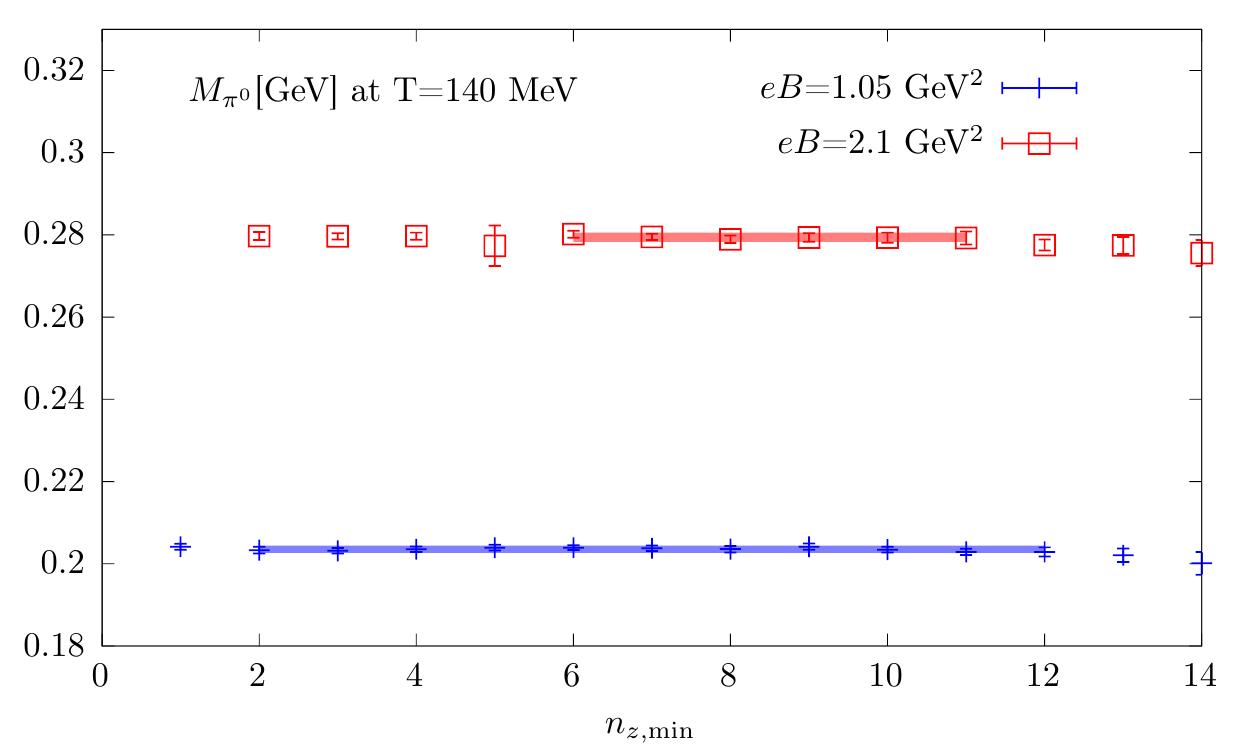}
	\includegraphics[width=0.328\textwidth]{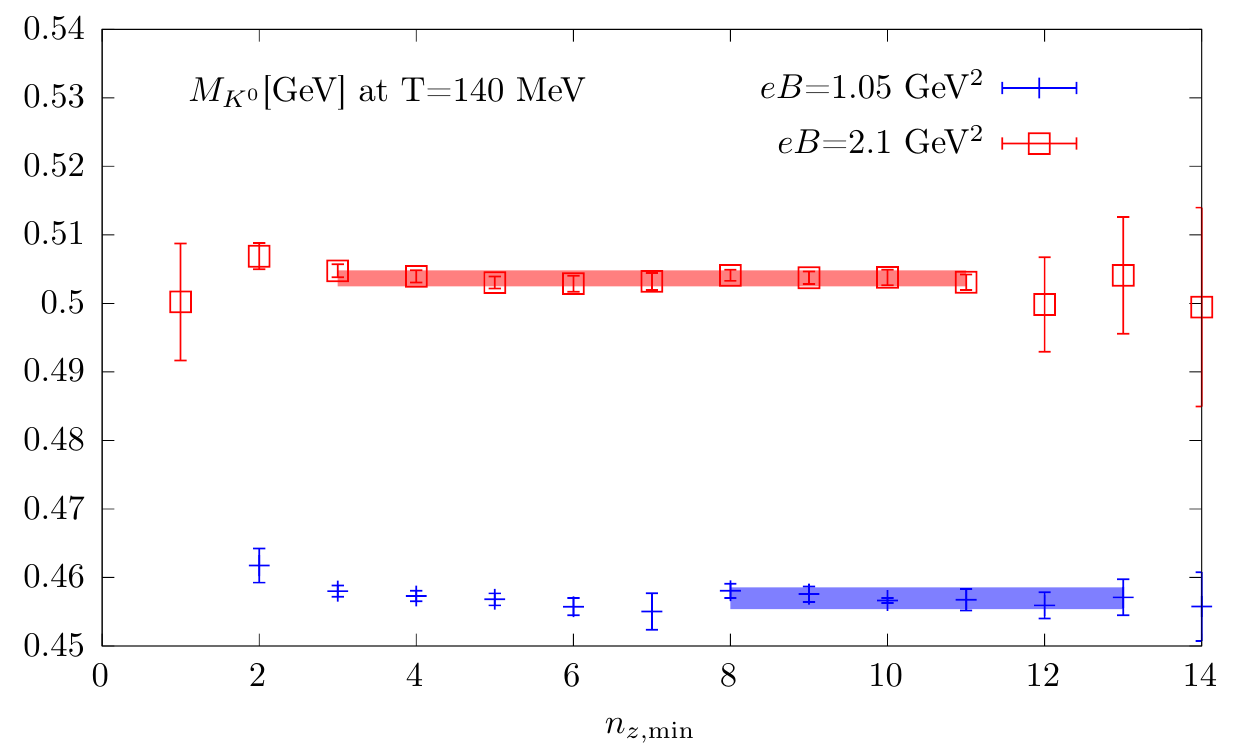} 
	\includegraphics[width=0.328\textwidth]{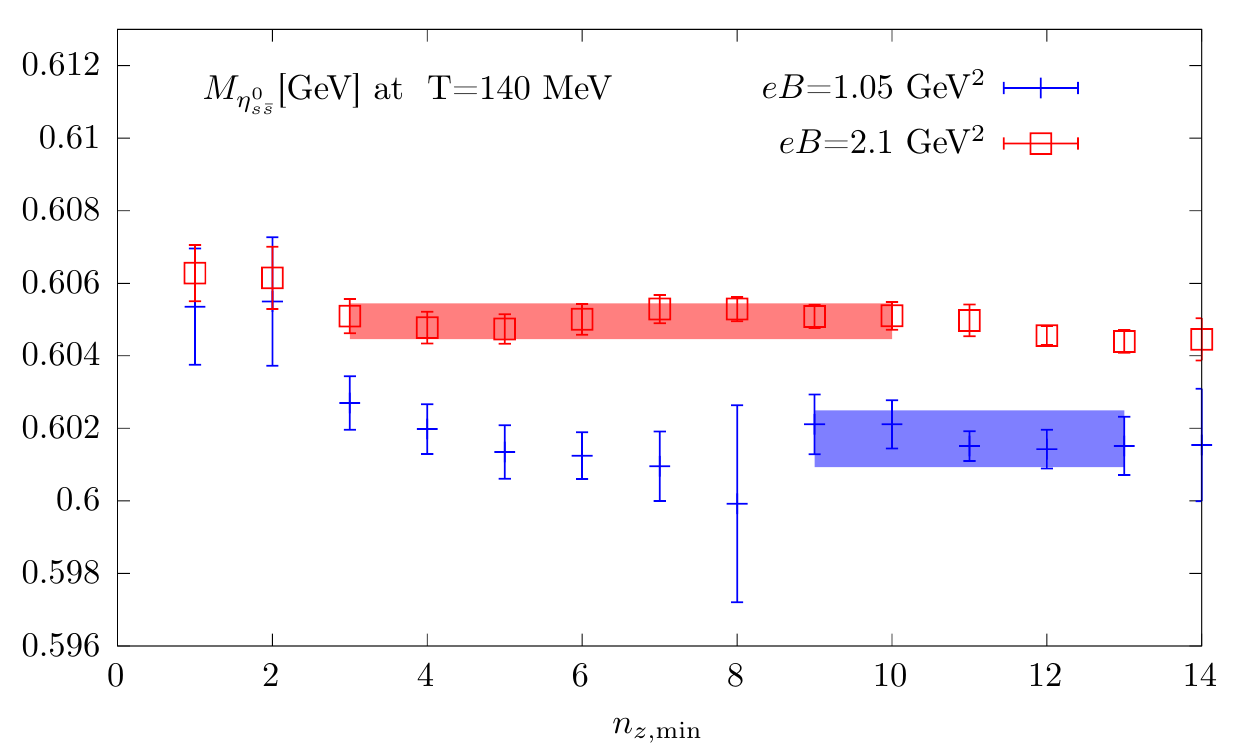} 
	\caption{Extraction of screening masses of $\pi^0$ (left), $K^0$ (middle) and $\eta^0_{s\bar{s}}$ (right) at $T$~=~140 MeV  with $N_b$~=~20 and 40. The fit interval is [$n_{z,{\rm min}},N_s/2$]. The bands show the final mean values and errors of the screening masses based on the AICc selected plateaus.}
	\label{fig:app_plateau}
\end{figure*}

\end{widetext}

%%------------------------------------------------------------------------------------
%       references
%------------------------------------------------------------------------------------
\bibliographystyle{JHEP}
\bibliography{WIextB}

%\begin{thebibliography}{99}
%
%
%
%\end{thebibliography}
%------------------------------------------------------------------------------------
%       end 
%------------------------------------------------------------------------------------
\end{document}